\documentclass[a4paper,useAMS,usenatbib]{mn2e}

\usepackage{amsmath}
\usepackage{amssymb,amsfonts}
\usepackage{graphicx,epsfig,graphics}
\usepackage{color}
\usepackage{supertabular}
\usepackage{subfigure}
\newcommand{\ba}{\mbox{\boldmath{$\alpha$}}}

\newcommand{\bg}{\mbox{\boldmath{$\gamma$}}}
\newcommand{\bk}{\mbox{\boldmath{$\kappa$}}}
\newcommand{\bd}{\mbox{\boldmath{$\delta$}}}

\begin{document}

\label{firstpage}
\title[Weak lensing reconstructions in 2D \& 3D]{Weak lensing reconstructions in 2D \& 3D: implications for
  cluster studies} \author[A. Leonard, F. Lanusse, J.-L. Starck]
      {Adrienne Leonard$^{1}$\thanks{Email:
          adrienne.leonard@ucl.ac.uk}, Fran\c{c}ois Lanusse$^2$, Jean-Luc
        Starck$^2$\\$^1$Department of Physics and Astronomy,
        University College London, Gower Place, London WC1E 6BT,
        U.K.\\$^2$Laboratoire AIM, UMR CEA-CNRS-Paris 7, Irfu, Service
        d'Astrophysique, CEA Saclay, F-91191 GIF-SUR-YVETTE CEDEX,
        France.}

\date{}
\maketitle

\pagerange{\pageref{firstpage}--\pageref{lastpage}} \pubyear{2014}

\begin{abstract}
We compare the efficiency with which 2D and 3D weak lensing mass
mapping techniques are able to detect clusters of galaxies using two
state-of-the-art mass reconstruction techniques: MRLens in 2D and
GLIMPSE in 3D. We simulate otherwise-empty cluster fields for 96
different virial mass-redshift combinations spanning the ranges
$3\times10^{13}h^{-1}M_\odot \le M_{vir}\le 10^{15}h^{-1}M_\odot$ and
$0.05 \le z_{\rm cl} \le 0.75$, and for each generate 1000
realisations of noisy shear data in 2D and 3D. For each field, we then
compute the cluster (false) detection rate as the mean number of
cluster (false) detections per reconstruction over the sample of 1000
reconstructions. We show that both MRLens and GLIMPSE
  are effective tools for the detection of clusters from weak lensing
  measurements, and provide comparable quality reconstructions at low
  redshift. At high redshift, GLIMPSE reconstructions offer increased
  sensitivity in the detection of clusters, yielding cluster detection
  rates up to a factor of $\sim 10\times$ that seen in 2D
  reconstructions using MRLens. We conclude that 3D mass mapping
techniques are more efficient for the detection of clusters of
galaxies in weak lensing surveys than 2D methods, particularly since 3D reconstructions yield unbiased estimators of both the mass and redshift of the detected clusters directly.

\end{abstract}


\section{Motivation}

Current and upcoming large surveys such as the Dark Energy Survey
\citep[DES, ][]{des}, Euclid \citep{euclid}, and the Large Synoptic
Survey Telescope \citep[LSST, ][]{lsst1,lsst2} will soon provide a
large volume of high-quality weak lensing data covering a significant
fraction of the sky up to redshifts $z\sim2$. While the primary weak
lensing science aims of these projects centre around the use of
two-point statistics to constrain the cosmological parameters, there
is nonetheless significant interest in reconstructing the mass
distribution inferred by the lensing shear field. Such maps are
obtained essentially for free in weak lensing surveys, as mass
reconstruction algorithms require no additional data inputs to
two-point shear statistics.

In addition to providing a valuable method for visualising the
distribution of structure in the Universe, such maps can help to
facilitate the computation of higher-order lensing statistics, in
addition to providing a straightforward method to detect large
structures such as clusters of galaxies. This can provide substantial
insights into the nature and evolution of nonlinear structure in the
Universe, and can help to break the degeneracy seen between
cosmological parameters in constraints obtained from two-point shear
statistics \citep[e.g. ][]{bernardeau97, Schneider98, TakadaJain2003,
  TakadaJain2004, JBJ04, ks05, pires09, berge10, DH10, pls12}.  Until
very recently, weak lensing mass reconstruction efforts focused on
estimating the two-dimensional projected mass density, the convergence
$\kappa$, integrated along the line of sight. With the advent of
accurate photometric redshift estimation methods and high-quality
data, it has now become possible to consider reconstructing the full
three-dimensional density field \citep{sth09, simonetal11, simon12,
  vanderplasetal11}. One advantage to this approach is that, by
directly reconstructing the density field, we might expect to be able
to directly estimate the masses of haloes detected without relying on
assumptions about the dynamical state of the halo. In a recent
publication, \cite{LLS14a} demonstrated a sparsity-based approach to
3D lensing mass reconstructions \citep[see also ][]{LDS12} that is
able not only to detect massive clusters of galaxies, but also to
estimate their masses and redshifts in an unbiased way. This
represents the first method able to reconstruct the density field from
weak lensing measurements sufficiently accurately to allow direct
estimates of the properties of dark matter haloes detected.

Such reconstructions, if sufficiently sensitive, might therefore be
able to augment and complement optical cluster studies that aim to
identify and weigh clusters of galaxies by considering the
distribution and/or dynamics of the cluster members
\citep[e.g. ][]{girardi98, lokas03, lokas06, mamon10,
  falco14}. Clusters of galaxies are a very important cosmological
probe. Studying the evolution of the cluster mass function
\citep{Rosati02,Voit05}, allows us to constrain both the amplitude of
the power spectrum at the cluster scale and the linear growth rate of
density perturbations. In addition, the clustering properties of the
large-scale distribution of clusters provides direct information on
the shape and amplitude of the underlying dark matter distribution and
power spectrum \citep{Borgani01,Moscardini01}.

In this paper, we focus on two state-of-the-art weak lensing mass reconstruction techniques: GLIMPSE in 3D \citep{LLS14a} and MRLens in 2D \citep{spr06}. The methodology underpinning the MRLens algorithm is qualitatively similar to that of GLIMPSE: both methods use wavelets, and impose sparsity to regularise the reconstructions and to control noise peaks. Other reconstruction methods in 2D could, of course, be chosen. An obvious example is the aperture mass statistic, which is frequently used for 2D mass mapping and peak detection. However, it was demonstrated in \cite{LPS12} that the aperture mass statistic is formally identical to the wavelet transform of the convergence map, with the added advantages that 
\begin{enumerate}
\item the wavelet transform considers several angular scales simultaneously, while application of the aperture mass statistic requires the choice of a scale radius, 
\item the wavelet transform is between 10 to 1000 times faster depending on the scale \citep{LPS12}, and 
\item the wavelet function intrinsically has some desirable properties, such as regularity, and compact support, and as a compensated filter it does not require truncation, unlike some aperture mass filters \citep[see][and references therein]{LPS12, pls12}.
\end{enumerate} 
It was further shown in \cite{pls12} that MRLens produces 2D reconstructions that are more robust to noise peaks than the aperture mass statistic, and are therefore more useful for cosmological studies such as those involving peak counts. It was aso shown that MRLens outperforms other convergence reconstruction methods such as  inverse Wiener filtering \citep{spr06}. For these reasons, and given the similarity to GLIMPSE in its wavelet- and sparsity-based approach, MRLens seemed the natural choice of algorithm for such a comparison.

We examine the ability of weak lensing mass
reconstruction techniques in two- and three-dimensions to detect
clusters as a function of the cluster mass and redshift. Using
simulated data, we consider 96 samples in virial mass and redshift
$[M_{vir,j},z_j]$, $j=\{1..96\}$, and for each generate 1000 noise realisations. We
generate reconstructions of these noisy fields using MRLens and GLIMPSE, and consider both the probability of detecting
a real cluster of a given mass and redshift (the true detection rate), and the probability of finding a
peak in a given field that arises 
due to noise (the false detection rate). Comparing the true detection rate in 2D and 3D reconstructions at the same peak signal-to-noise threshold allows us to probe any intrinsic differences in the lensing signal-to-noise when the full 3D information is retained versus when this information is projected along the line of sight. However, naively we might expect the false detection rate to scale with the number of pixels in the reconstruction, which is always larger in 3D than in 2D. The ideal reconstruction algorithm will maximise the true detection rate and minimise the false detection rate, and we examine how these compare between the two algorithms, and scale as a function of user-specified algorithm parameters.

This paper is organised as follows: In \S~\ref{sec:methods}, we
provide a brief overview of weak lensing mass reconstructions, and
outline the MRLens and GLIMPSE algorithms. In \S~\ref{sec:results} we
describe the suite of cluster simulations, and present the results of
our study as described above. We conclude with a brief discussion of
our results and their implications for cluster and weak lensing
surveys in \S~\ref{sec:discussion}.

\section{Mass Reconstruction Methods}
\label{sec:methods}

In what follows, we assume that the vector 
\begin{equation}
\mathbf{\gamma}^{(i)} = \left(\epsilon_1^{(i)} \epsilon_2^{(i)} ... \epsilon_{N_g}^{(i)}\right)^t
\label{eq:ell}
\end{equation}
contains the complex ellipticities of galaxies in a tomographic weak lensing survey binned on a grid with $N_g$ grid points. Sources are binned together on the angular grid only if they belong to the same redshift bin, which is denoted by the superscript $^{(i)}$ in equation \eqref{eq:ell} above. We can express the full three-dimensional weak lensing information as the vector $\bmath{\gamma}\equiv(\bmath{\gamma}^{(1)},\bmath{\gamma}^{(2)}, ... , \bmath{\gamma}^{(N_z)})$, where $N_z$ is the number of redshift bins in the survey.

The aim of mass reconstruction methods is to relate the measured shear
$\bmath{\gamma}$ to the underlying density distribution. Typically, in 2D, we aim
to recover the projected, dimensionless surface density (the
convergence) $\bmath{\kappa}^{(i)}$ from the tomographic shear $\bmath{\gamma}^{(i)}$, while in 3D we seek the dimensionless 3D matter
overdensity $\bmath{\delta}$. We can relate the tomographic shear $\bmath{\gamma}^{(i)}$ to the convergence $\bmath{\kappa}^{(i)}$ of the lens by
\begin{equation}
\boldsymbol{\gamma}^{(i)}(\theta) = \frac{1}{\pi}\int d^2\theta^\prime \mathcal{D}(\theta-\theta^\prime)
\boldsymbol{\kappa}^{(i)}(\theta^\prime)\ ,
\label{eq:2dlenseq}
\end{equation}
where $\theta = \theta_1+\mathrm{i}\theta_2$ represents the angular
coordinate on the sky, represented in complex notation,
\begin{equation}
\mathcal{D}(\theta) = \frac{1}{(\theta^\ast)^2}\ ,
\end{equation}
and the asterisk $^\ast$ represents complex conjugation. 

Equation \eqref{eq:2dlenseq} represents a linear mapping between the
shear and the convergence, and can be expressed in matrix
notation as $\bg^{(i)} = \mathbf{P}_{\gamma\kappa}\bk^{(i)} + \boldsymbol{n}_\gamma^{(i)}$, where $\bmath{n}_\gamma^{(i)}$ is a vector of intrinsic source ellipticities and the linear transform $\mathbf{P}_{\gamma\kappa}$ is, for $\ell\ne m$:
\begin{equation}
[\mathbf{P}_{\gamma\kappa}]_{\ell m} = -\frac{A}{\pi}\frac{1}{[\theta_{\ell m}^\ast]^2}\ ,
\end{equation}
where $A$ represents the solid angle of the angular grid pixels, and $\ell$ and $m$ give the $x$ and $y$ coordinates of the pixel on the grid. When $ \ell = m$, we set $\mathbf{P}_{\gamma\kappa}=0$. 

The underlying convergence, though representing a two-dimensional projected density, is actually itself a three-dimensional field: at each point $\boldsymbol{r}$ in three-dimensional space, we can compute a convergence $\bk(\boldsymbol{r})$, by considering the relationship of the lensing convergence to the underlying matter overdensity $\boldsymbol{\delta}(\boldsymbol{r})$. Note that we often parameterise the 3D position vector $\boldsymbol{r}$ in observations by an angular position $\theta$ and a redshift $z$. This matter overdensity, or density contrast, is defined as $\boldsymbol{\delta}(\theta,z)\equiv
(\boldsymbol{\rho}(\theta,z)-\overline{\rho(z)})/\overline{\rho(z)}$, where $\boldsymbol{\rho}(\theta,z) $ is the density at angular position $\theta$ and redshift $z$ and $\overline{\rho}(z)$ is the mean matter density at redshift $z$, and is related to the convergence by
\begin{eqnarray}
\small \bk(\theta,z) = \frac{3H_0^2\Omega_M}{2c^2} \int_0^w dw^\prime
\frac{f_K(w^\prime)
  f_K(w-w^\prime)}{f_K(w)}\frac{\boldsymbol{\delta}[f_K(w^\prime)\theta,w^\prime]}{a(w^\prime)}\ ,
  \label{eq:kapconv}
\end{eqnarray}\normalsize
where $H_0$ is the hubble parameter, $\Omega_M$ is the matter density
parameter, $c$ is the speed of light, $a(w)$ is the scale parameter
evaluated at comoving distance $w$, and
\begin{equation}
  f_K(w) = \begin{cases} K^{-1/2}\sin(K^{1/2}w), & K>0 \\ w, & K=0
    \\ (-K)^{-1/2}{\rm sinh}([-K]^{1/2}w) & K<0 \end{cases}\ ,
\end{equation}
gives the comoving angular diameter distance as a function of the
comoving distance and the curvature, $K$, of the Universe. $f_K(w)\theta$ in the argument of $\bmath{\delta}$ in equation \eqref{eq:kapconv} therefore gives the transverse comoving distance.  If we now consider that we have a tomographic measurement of the convergence, $\bmath{\kappa} = (\bmath{\kappa}^{(1)},\bmath{\kappa}^{(2)},...,\bmath{\kappa}^{(N_z)})$, obtained from the tomographic shear $\bg$, the above relationship can also be expressed in matrix notation as $\bk =
\mathbf{Q}\bd$, where $\mathbf{Q}$ represents the line of sight convolution operation in equation \eqref{eq:kapconv}, and is given in full in \cite{sth09} and \cite{LLS14a}, and $\bd$ now represents a binned map of the density contrast\footnote{Note that the redshift binning in $\bd$ does not have to be the same as that of the convergence and shear; i.e. the matrix $\mathbf{Q}$ can be rectangular rather than square.}. For a given set of shear measurements
$\bmath{\gamma}$, we therefore can write
\begin{eqnarray}
{\bg} &=& \mathbf{P}_{\gamma\kappa}\bk + \boldsymbol{n}_\gamma\ ,\label{eq:gamkap}\\
&=& \mathbf{P}_{\gamma\kappa}\mathbf{Q}\bd + \boldsymbol{n}_\gamma\ ,\label{eq:gamdelt}
\end{eqnarray}
where, following the conventions used previously,  $\bmath{n}_\gamma = (\bmath{n}_\gamma^{(1)},\bmath{n}_\gamma^{(2)},...,\bmath{n}_\gamma^{(N_z)})$ represents the error in the shear measurements. While there
are various sources of error in shear measurements, for the purposes
of this paper we consider only noise arising from the intrinsic shapes
of galaxies, which is typically taken to be a Gaussian distribution of
width $\sigma_\gamma\sim 0.2-0.3$.

The task of mass reconstruction methods is the inversion of either
equation \eqref{eq:gamkap} in 2D or equation \eqref{eq:gamdelt} in
3D. Note that the formalism presented above assumes the absence of intrinsic alignments, and this assumption is propagated throughout the analysis. A full assessment of the impact of intrinsic alignments on the reconstruction quality is the subject of future work.

We outline below the methods underlying two state-of-the-art mass
reconstruction techniques: MRLens \citep{spr06} in 2D and GLIMPSE
\citep{LLS14a} in 3D.

\subsection{2D mapping with MRLens}

In 2D, our aim is to solve Equation \eqref{eq:gamkap} for each tomographic redshift bin and estimate
the convergence $\bk^{(i)}$ from noisy shear measurements. In the Fourier
plane, the convergence can be computed from the shear components as \citep{KS93}
\begin{equation}
\tilde{\bk}^{(i)}(\mathbf{k}) = \tilde P_1(\mathbf{k}) \tilde{\bg_1}^{(i)} + \tilde P_2(\mathbf{k})  \tilde{\bg_2}^{(i)},
\label{eq:KS_inversion}
\end{equation}
where the hat symbol denotes Fourier transforms and we have defined $k^2 \equiv k_1^2 + k_2^2$ and
\begin{eqnarray}
\tilde{P_1}(\mathbf k) & = & \frac{k_1^2 - k_2^2}{k^2} \nonumber \\
\tilde{P_2}(\mathbf k) & = & \frac{2 k_1 k_2}{k^2},
\end{eqnarray}
with $\tilde{P_1}(\mathbf{k}) \equiv 0$ when $k_1^2 = k_2^2$, and
$\tilde{P_2}(\mathbf{k}) \equiv 0$ when $k_1 = 0$ or $k_2 = 0$. These
conditions on $\tilde{P_1}$ and $\tilde{P_2}$ when $k_1=k_2=0$ correspond
to an indetermination of the mean value of the convergence field
$\kappa$, known as the mass-sheet degeneracy. Applying an inverse
Fourier transform on $\tilde{\kappa}$ estimated by equation
\eqref{eq:KS_inversion} yields an estimate of the convergence. As this
estimate is typically very noisy, some type of filtering or
regularisation is usally included in a reconstruction algorithm in
order to minimise the contaminating effect of noise on the resulting
reconstruction. Gaussian and Wiener filtering are common choices, but
may not be optimal, particularly since Wiener filtering is only optimal when both the signal and noise are Gaussian-distributed, and their power spectra are known. 

MRLens is a Multi-Resolution Entropy filtering method using a wavelet
based prior for the entropy to regularise the solution. For full details of the MRLens algorithm, the reader is referred to \cite{spr06}; briefly, this method assumes that the reconstructed convergence can be represented \textit{sparsely} when projected onto a well-chosen dictionary or set of functions (basis or frame). By this we mean that most of the information content can be captured by a small number of coefficients in the chosen dictionary or frame. For example, a periodic signal would be sparse in the Fourier domain, as it would be completely represented by a small number of Fourier coefficients. In the present case, we choose wavelets. 

In the case of Gaussian noise, one can construct a \textit{multiresolution support} in the wavelet domain by identifying as ``significant" those wavelet coefficients $w_{j,\ell,m}$ that verify $|w_{j,\ell,m}| > k \sigma_j$, where $\sigma_j$ is the noise standard deviation at a particular wavelet scale $j$, and the indices $[\ell, m]$ again denote the pixel location in the 2D convergence map\footnote{We note that MRLens offers a second method to determine significant wavelet coefficients based on the False Discovery Rate \citep[][and references therein]{spr06}, which may improve the quality of the reconstructions by limiting the ratio of false detections to true detections to a user-specified value. However, for ease of comparison with GLIMPSE, which defines significant coefficients based on their SNR, we chose to use the $k\sigma$ method in this work.}. The noise standard deviation can be estimated directly from the noisy map, and we typically choose the threshold $k$ to be between 3 and 5. Coefficients that are deemed thus to be significant are retained in the reconstruction, while an entropy based regularisation is applied to those coefficients that are not significant. An estimate of the convergence is built iteratively, through application of this regularisation on the residual between the data and the current convergence estimate.

\subsection{3D mapping with GLIMPSE}

The full 3D problem presents an additional difficulty compared to the
2D problem as we are trying to de-project the structures along the
line of sight. However, the radial lensing operator $\mathbf{Q}$ is
singular and leads to an ill-posed inverse problem, which does not
accept a single, stable, solution even in the absence of noise. A
robust method to address this very general class of problems is sparse
regularisation, as in the 2D case, above. If the signal to recover is assumed to be sparse (only
a small number of coefficients are non-zero) in an adapted dictionary,
then a robust estimate of the original signal can be recovered by
solving a optimisation problem with a penalty on the $\ell_1$ norm of
the coefficients of the signal in the chosen dictionary \citep{smf10}. GLIMPSE
implements this approach in the context of the 3D mapping problem and
recovers an estimate of the density contrast $\delta$ as the solution
of the following minimisation problem:
\begin{equation}
\centering
\min_{\boldsymbol{\alpha}} \frac{1}{2} \parallel \bg - \mathbf{P}_{\gamma\kappa} \mathbf{Q} \mathbf{\Phi} \ba \parallel^{2}_{\mathbf{\Sigma}} + \lambda \parallel \ba \parallel_{1}\ ,
\label{eq:WLLagrangian}
\end{equation}
where $\ba$ are the coefficients of the estimate of the density contrast $\bd$
in an appropriate dictionary $\mathbf{\Phi}$ (i.e. $\bd = \mathbf{\Phi}\ba$), $\lambda$ is a parameter
tuning the sparsity constraint and, as before, $\mathbf{\Sigma}$ is the covariance matrix of the noise, assumed to be Gaussian. The first term in equation \eqref{eq:WLLagrangian} is called the \textit{data fidelity} term, and represents the $\ell_2$ distance between the data and the current estimate of the solution, weighted by the noise covariance, while the second term represents a sparsity penalty, i.e. the equation above is minimised when the $\ell_1$ norm of the vector of coefficients is as small as possible whilst being as close as possible to the input data.

GLIMPSE uses a 2D-1D dictionary for
$\mathbf{\Phi}$ composed of isotropic wavelets in the 2D angular domain and
Dirac delta functions in the radial dimension. Such a representation
is well adapted to the expected density contrast of a typical dark
matter halo, which is always contained within a single redshift bin
given the resolutions we are able to attain in the radial dimension.

GLIMPSE solves this optimisation problem using a variant of the Fast
Iterative Soft Thresholding Algorithm (FISTA, \cite{FISTA2009}). This
algorithm relies on two steps, first the coefficients $\alpha$ of the
solution are updated according to a gradient descent of the data
fidelity term, then the coefficients are thresholded,
setting to zero coefficients of amplitude lower than $\lambda$. This imposes the sparsity constraint on the solution. As with MRLens, GLIMPSE sets the threshold level to
$\lambda_{j,n} = k \sigma_{j,n}$ where $\sigma_{j,n}$ is the noise standard deviation at wavelet scale $j$ and redshift bin $n$. With this choice of parameter, at each iteration only the significant
coefficients (i.e. of amplitude above $k \sigma_{j,n}$) are allowed to enter
the solution. The only parameter to set for the GLIMPSE algorithm is
the significance threshold $k$; as with MRLens, this is usually taken
to be between 3 and 5.

\section{Tests with cluster simulations}
\label{sec:results}

\begin{figure}
\hskip5ex
\includegraphics[width=0.45\textwidth]{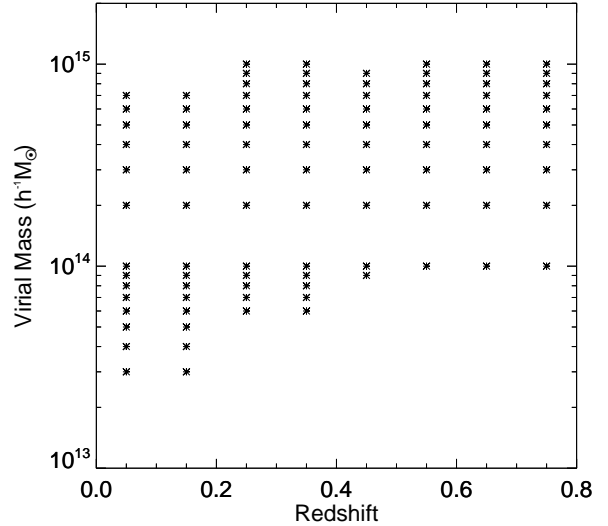}
\caption{This figure shows the virial mass and redshift for the 96
  clusters forming this study. \label{fg:sims}}
\end{figure}

In order to test the effectiveness of the two mapping techniques at
detecting and reconstructing clusters of galaxies, we simulated 96
clusters of galaxies at a range of masses and redshifts as in
\cite{LLS14a}; Figure \ref{fg:sims} shows the points sampled in virial
mass-redshift space. Each cluster halo follows an NFW profile
\citep{NFW97} with a concentration parameter computed from the virial
mass and redshift using the parameterisation given in
\cite{couponetal12}.

From the NFW density profile, we computed the corresponding shear
signal, which is derived analytically in \cite{TakadaJain2003}. The
computations required to simulate the halos were performed making
extensive use of the NICAEA software
package\footnote{http://www2.iap.fr/users/kilbinge/nicaea/}, using a
flat $\Lambda$CDM cosmology with $\Omega_{\rm M} =
0.264,\ \Omega_\Lambda = 0.736$, and $H_0 = 71$km/s/Mpc. 

Each cluster was simulated in an otherwise empty field with an angular
pixel size of $1^\prime\times 1^\prime$. We considered our shear
information to come from galaxies following a uniform angular
distribution on the sky, with a redshift distribution given
by 
\begin{equation} n(z) = z^{\alpha} \exp\left( -
  \left[\frac{z}{z_0}\right]^{\beta} \right)\ ,
\end{equation}
where we take $z_0 = 1/1.4$, giving a median redshift $z_{\rm med} = 1$, $\alpha = 2$ and $\beta = 1.5$ for a Euclid-like survey \citep{tayloretal07, kitchingetal11}. We assumed
30 galaxies per square arcminute, and that the intrinsic galaxy
ellipticity follows a Gaussian distribution of width
$\sigma_\varepsilon = 0.25$. In their current forms, both MRLens and GLIMPSE take as inputs the shear, rather than the reduced shear ($g = \gamma/(1-\kappa)$), and so our simulations produce noisy shear fields, rather than reduced shear fields. This is a simplification, included simply for ease of computation of our simulations, but which may lead to a biased reconstruction if applied to real data. However, \cite{Jullo2014} describe an iterative scheme to account for the reduced shear that converged after three iterations and de-biases the reconstruction. Such an iterative process can be implemented with both MRLens and GLIMPSE in applications to real data to account for reduced-shear. This will be demonstrated for GLIMPSE in future work.

Finally, we assumed Gaussian photometric
redshift errors with a standard deviation given by $\sigma_z =
0.05(1+z)$ and no bias.

The 2D projected shear map was computed by integrating the lensing
signal along each line of sight, while the 3D shear map was computed
using 30 tomographic bins of variable width $\Delta z$, such that each
tomographic bin contained the same number of galaxies. The bins were chosen to be sequential in photo-z space. Given the photometric redshift errors, such a choice means that the true redshift distributions for each bin overlap \citep[see, e.g., figure 2 of][]{LLS14a}. Each map was
generated with $64 \times 64$ angular pixels, each of width $\delta
\theta = 1$ arcmin. In order to properly asses the statistics of the
reconstructions, for each cluster field, we computed 1000 noisy
realisations of the data.

\subsection{Algorithm configurations}
\label{subsec:algoconfig}

In order to compare the two methods in a fair way, we used algorithm
configurations that were as similar as possible. Both MRLens and
GLIMPSE were run using the same 2D wavelets \citep[the starlet,
][]{Starck07Starlet} using 6 wavelet scales in the decomposition. As
GLIMPSE applies a $k\sigma$ hard threshold in density space, we
configured MRLens to use $k\sigma$ thresholding, rather than the
default False Discovery Rate (FDR) thresholding. This is not the
optimal configuration for MRLens, as it may allow a higher fraction of
false peaks to appear in a given reconstruction, but was chosen in
order to provide a fair basis for comparison with GLIMPSE. The
threshold level was chosen to be at 4$\sigma$ for GLIMPSE. For MRLens,
we considered three different thresholds:
$3\sigma,\ 3.5\sigma,\ \mathrm{and }\ 4\sigma$.

In this paper, we are interested in both the probability of detecting
a real cluster of a given mass and redshift with a given denoising
threshold(the true detection rate), and the probability of finding a
peak in a given field that rises above the denoising threshold but is
due entirely to noise (the false detection rate). Comparing the true
detection rate between 2D reconstructions and 3D reconstructions at
the same denoising threshold will probe any intrinsic differences in
the lensing signal to noise when the signal is projected in 2D vs when
the full 3D information about the signal is retained. However, naively
we expect the false detection rate at a fixed denoising threshold to
scale with the number of pixels in the reconstruction, and therefore
naturally expect a higher level of contamination from spurious peaks
in a 3D reconstruction of a given lensing field than in the 2D
reconstruction of that field using the same denoising threshold.

A mass reconstruction algorithm will ultimately be judged both on its
ability to detect real clusters and its effectiveness at controlling
spurious peaks. There is always a trade-off to be had between
completeness, which may be increased by lowering the denoising
threshold, and purity, which is improved by raising the denoising
threshold. The setup of the present experiment does not allow us to
compute the expected purity we might obtain in a reconstruction of a
real weak lensing dataset. However, a comparison of the true detection
rates obtained by MRLens and GLIMPSE when the denoising thresholds
yield comparable false detection rates provides a more fair assessment
of the overall performance of the two methods in application to weak
lensing studies.

\subsection{Peak identification}

We used CLFIND \citep{williamsetal94}, which is a friends-of-friends
algorithm, to identify connected pixels above a given threshold
associated with a possible detection. The user is required to set a minimum value that a pixel can have in order to be considered as part of a detected structure. The aim is to identify the pixels associated with a detected peak out to large radii from the centre of that peak, so that we are able to accurately compute properties such as the density-weighted centroid for the detected peak.

Setting the detection threshold in 3D is fairly straightforward: the GLIMPSE
algorithm applies thresholding to the solution, meaning that
regions of the reconstruction in which there are deemed to be no
significant coefficients during the iterative reconstruction process
will be set to zero in the reconstruction. Therefore, any deviations
from zero may be considered a detection of a feature above the
noise. However, setting the CLFIND threshold to be zero can lead to blending of detected structures and an increased runtime for the algorithm, so to minimise any potential blending and
to optimise the runtime of the CLFIND algorithm in 3D, we set the
detection limit to be $\delta_{min} = 1$.

Defining a detection in 2D is somewhat more complicated. While MRLens
does apply hard thresholding during the iterative procedure, due to differences in the details of the algorithm -- specifically, that MRLens computes the multiresolution support and applies the entropy based regularisation to the residual, which is then added back on each iteration to build the solution -- there are no regions in the reconstructed
convergence maps that are identically zero. Furthermore, due to the
mass-sheet degeneracy, the overall normalisation of the reconstructed
convergence map may be somewhat arbitrary. In practice, this is usually constrained by either setting the mean convergence (over a sufficiently large field) to be zero or, in cluster fields, by setting the convergence to be zero sufficiently far away from the cluster. We apply no such boundary conditions in the 2D reconstructions presented here. However, we note that in 3D the density contrast is naturally physically constrained to be $\ge -1$ (i.e. the density is always positive), and GLIMPSE applies a positivity constraint to its estimate of the solution at each iteration.

\begin{figure}	
	\subfigure[The reconstructed convergence produced by MRLens, plotted on a logarithmic colour scale. \label{fg:kappa}]{\includegraphics[width=0.22\textwidth]{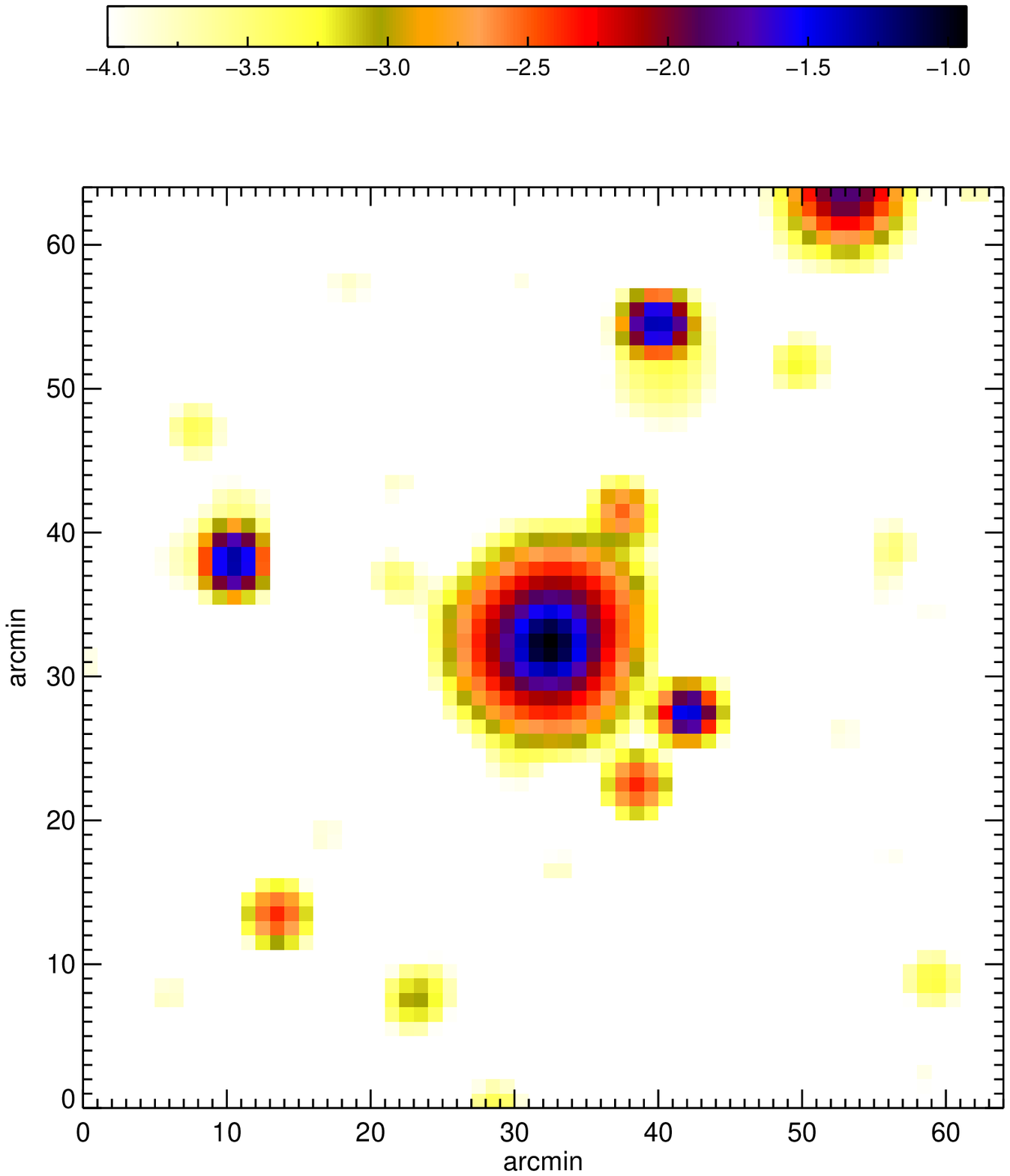}}\hfill
	\subfigure[The multiresolution support produced by MRLens, showing the pixel locations at which a significant wavelet coefficient was detected.\label{fg:support}]{\includegraphics[width=0.22\textwidth]{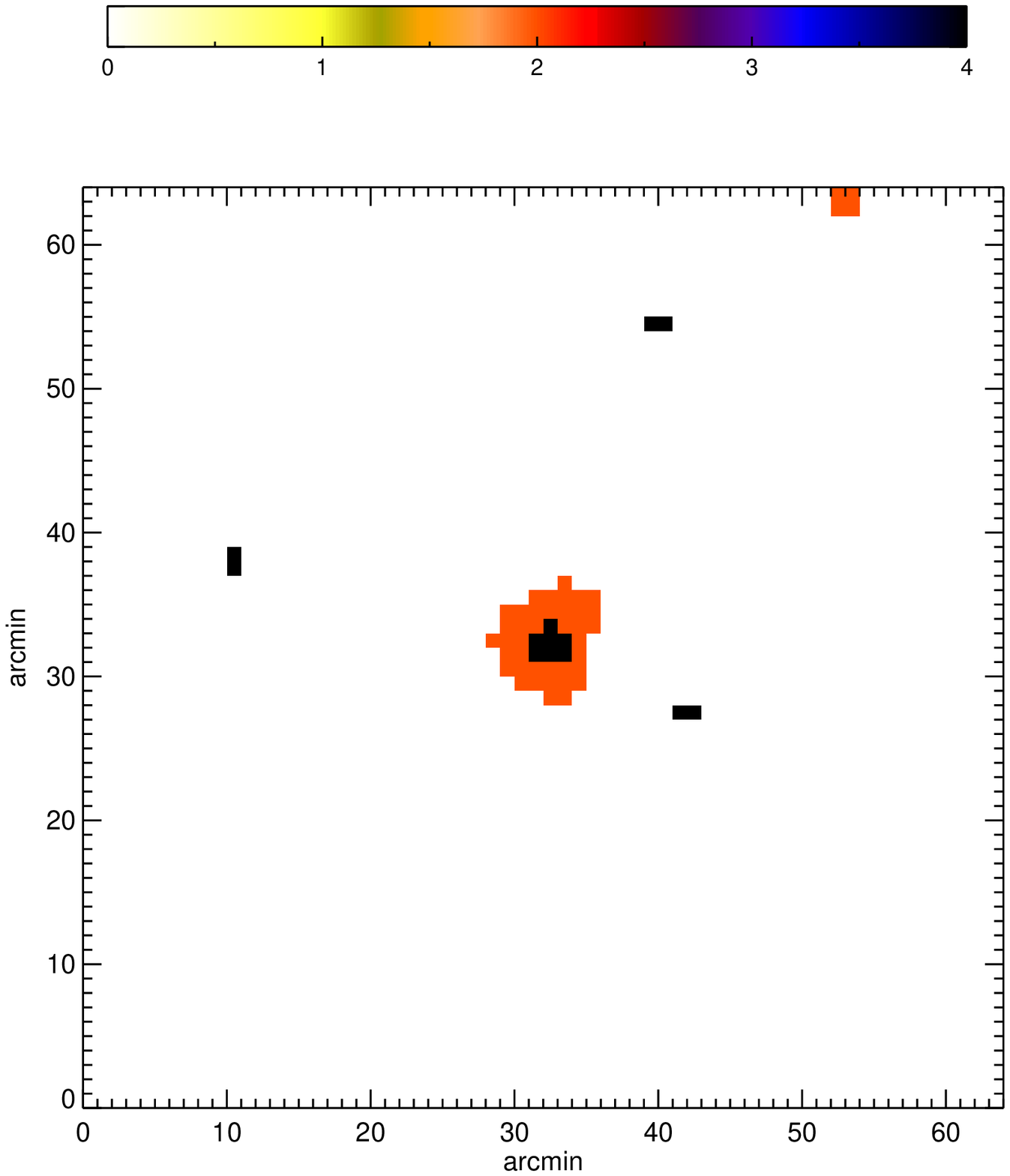}}\vfill
	\subfigure[Pixels not connected with regions of non-zero multiresolution support are masked and CLFIND run on the resulting image as illustrated above. We use here the same colour scheme as in figure \ref{fg:kappa} above.\label{fg:masked}]{\includegraphics[width=0.22\textwidth]{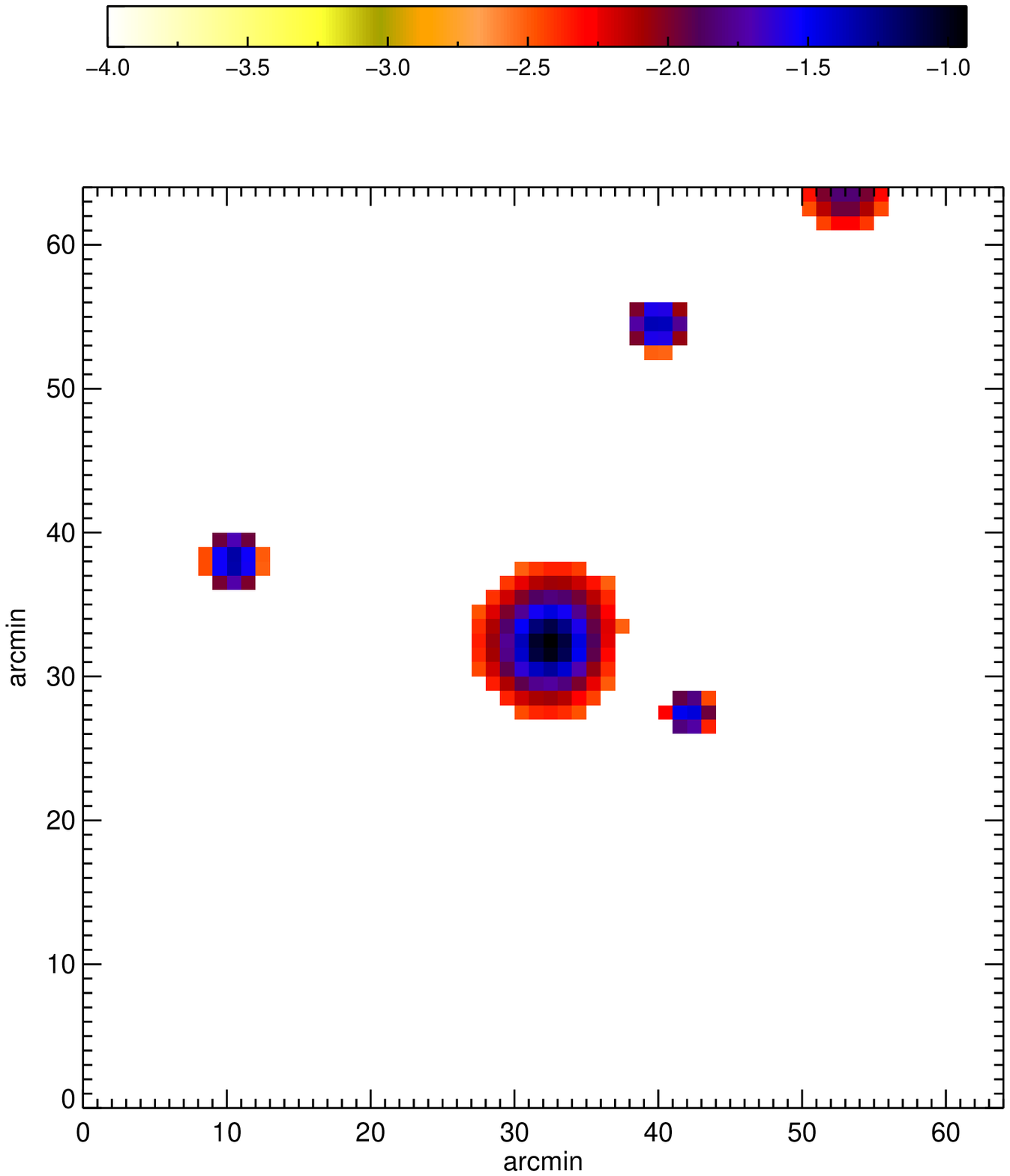}}\hfill
	\subfigure[The output from CLFIND: a mask identifying and labelling distinct structures in the reconstruction. The pixels associated with each clump are used to compute the convergence-weighted centroid of the peak.\label{fg:foundpeaks}]{\includegraphics[width=0.22\textwidth]{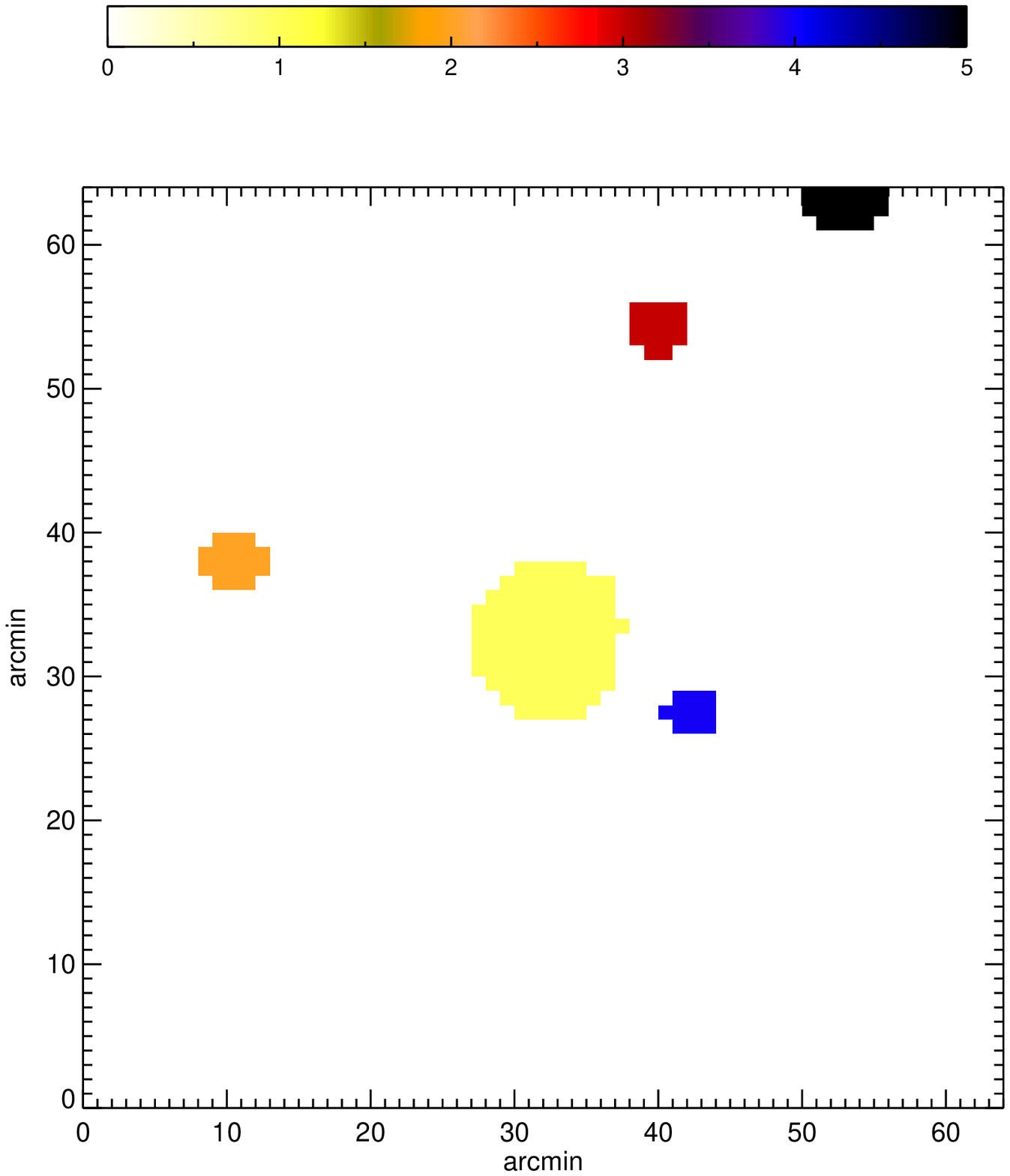}}
	\caption{The pipeline for detection of peaks in 2D MRLens reconstructions \label{fg:detectionsin2D}. The 3D pipeline is simpler, involving application of CLFIND directly on the 3D reconstructions, with no masking required.}
\end{figure}

To ensure that we all the structures identified in the multiresolution support and no others, we perform a multi-step detection procedure on all the 2D reconstructions (see figure \ref{fg:detectionsin2D} for illustrations of each step):
\begin{enumerate}
\item Identify all pixels in the convergence reconstruction (figure \ref{fg:kappa}) where the multiresolution support (figure \ref{fg:support}) is non-zero; find the minimum value of the convergence amongst these pixels, $\kappa_{min}$. We define $\kappa_{lim} = \min[3\times10^{-3},\kappa_{min}/2]$. 
\item Search for all pixels with values $\ge \kappa_{lim}$ that are connected to the pixels with non-zero multiresolution support. 
\item Mask all other pixels (figure \ref{fg:masked}).
\item Run CLFIND on the masked map to identify and separate the peaks detected in the map (figure \ref{fg:foundpeaks})
\end{enumerate}
Note that the limiting value of $3\times10^{-3}$ above was chosen as an effective trade-off between capturing as many pixels as possible associated with each peak detected and minimising blending and the CLFIND algorithm runtime. In cases where the convergence value at the locations of non-zero multiresolution support are close to this value, the limit is lowered in order to link a larger number of pixels to the detected peaks. This also ensures that peaks coinciding with non-zero multiresolution support are always detected. This does occasionally lead to blending, however CLFIND as the final step is able to isolate and separate blended peaks where two clear local maxima occur. 

%

We compute the location of each peak in 3D by computing the
density-weighted centroid $[x,y,z]$, and in 2D analogously by
computing the convergence-weighted centroid $[x,y]$ using those pixels identified by CLFIND.

\subsection{Peaks arising from the noise}
\label{subsec:noisepeaks}

Naively, we expect that when we consider the angular distribution of
all peaks detected in all 1000 reconstructions of a given cluster
field, we will find a strong tendency to detect a peak in the centre
of the field, where the cluster is located, and a roughly uniform
distribution of false peaks arising from the noise, at a much lower
amplitude. This is because false detections due to random noise are
expected to be uniformly randomly distributed, as the noise is uniform
and uncorrelated in the $x-y$ plane.

\begin{figure}
\includegraphics[width=0.24\textwidth]{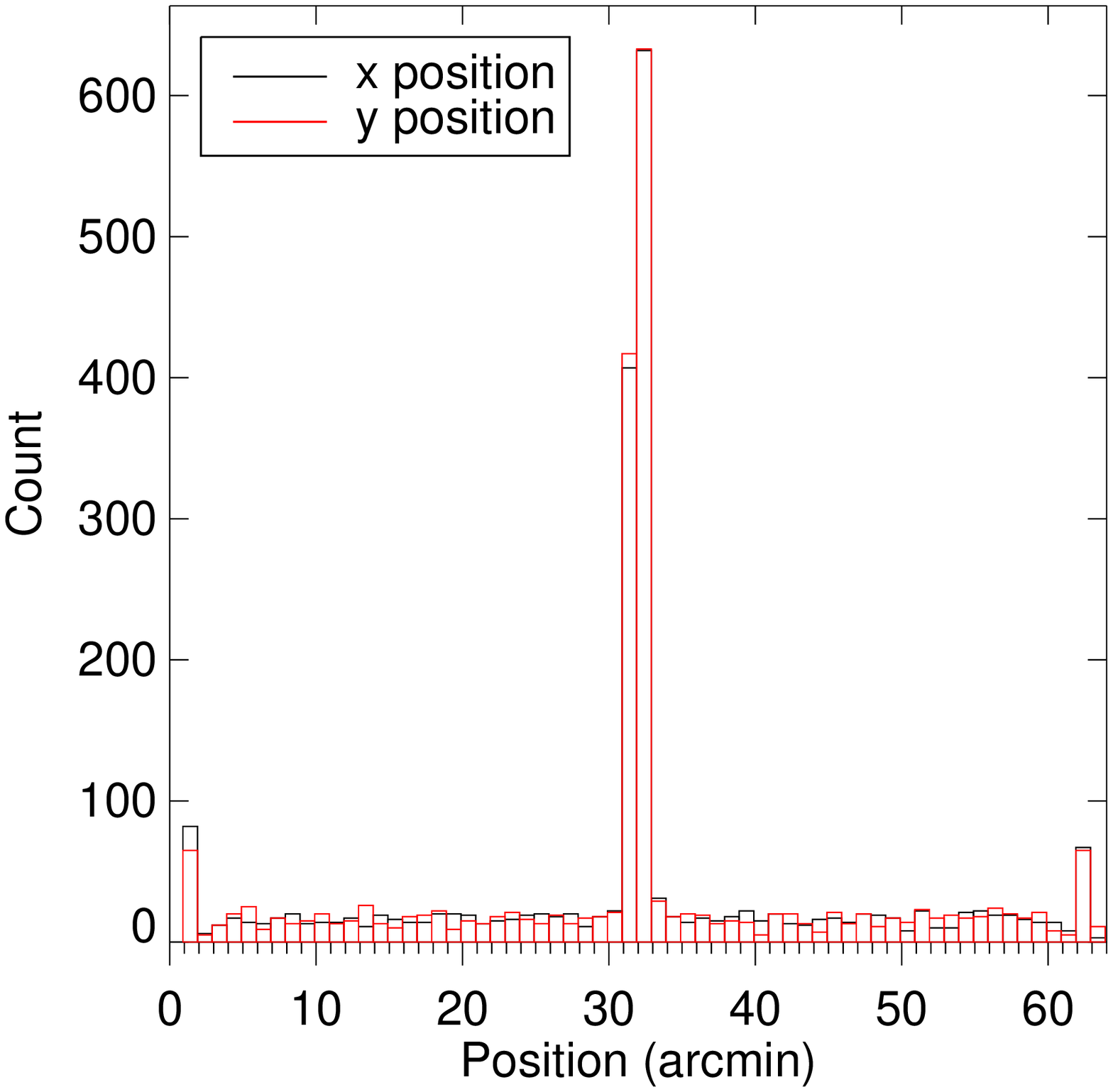}\includegraphics[width=0.24\textwidth]{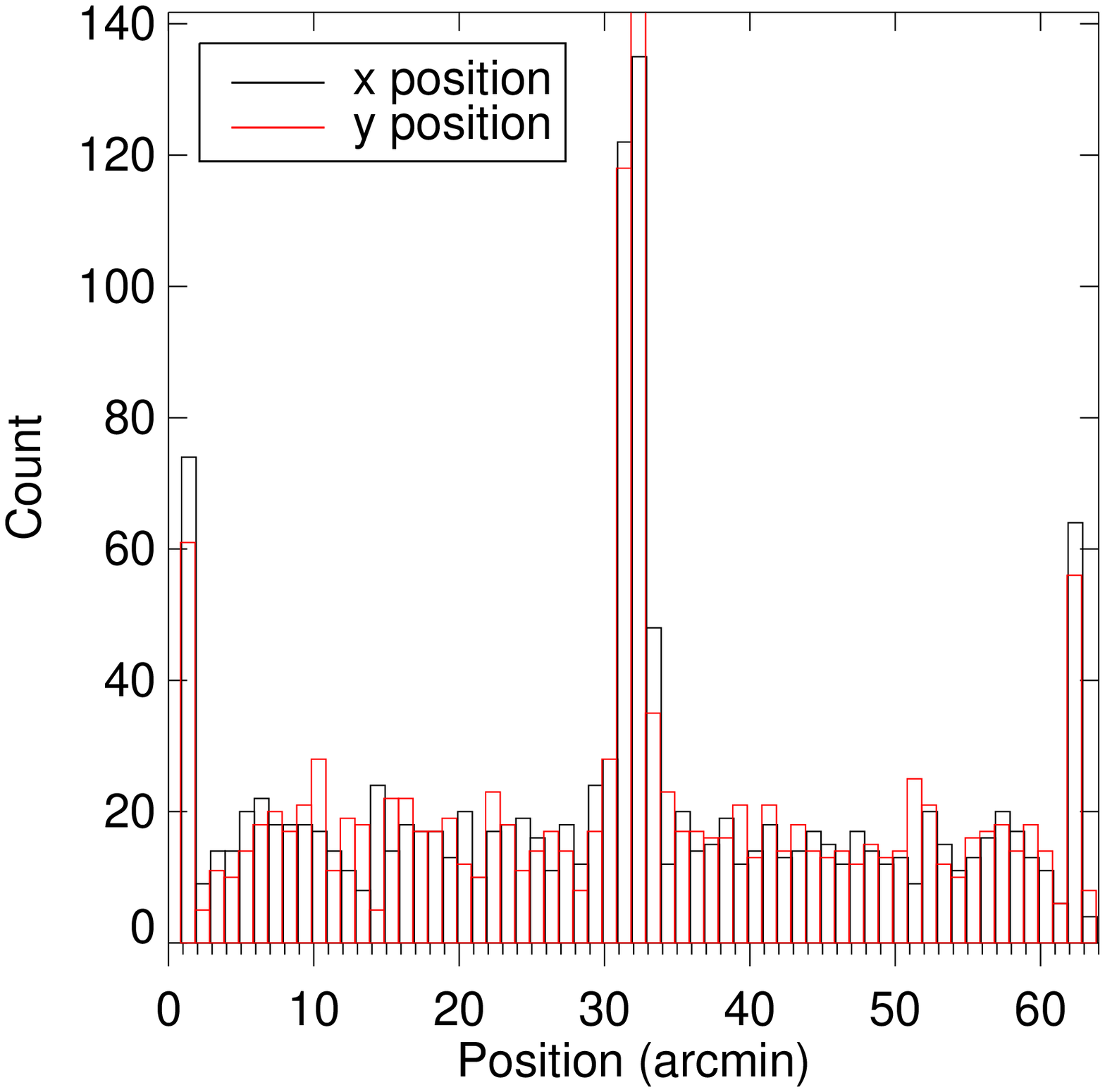}
\caption{This figure shows the distribution of the $x$ and $y$
  positions of peaks detected in 1000 GLIMPSE 3D reconstructions of a
  field containing a cluster of virial mass
  $7\times10^{14}h^{-1}M_\odot$ at a redshift of $z_{\rm cl}=0.35$ (\textit{left panel}) and the analogous plots for a field containing a smaller cluster of virial mass $9 \times 10^{13}h^{-1}M_\odot$ at the same redshift (\textit{right panel}). In both cases, the
  central peak represents detections of the cluster itself, while the
  approximately uniform background represents false detections. Some
  edge effects are apparent at extreme values of $x$ and
  $y$. \label{fg:xy3D}}
\end{figure}

In figure \ref{fg:xy3D} we plot a histogram of the $x$ and $y$
positions of all detected peaks in 1000 GLIMPSE 3D reconstructions of
a field containing a relatively high signal-to-noise cluster of virial mass $M_{vir} =
7\times10^{14}h^{-1}M_\odot$ located at a redshift of $z_{cl} =
0.35$, and an analogous plot for reconstructions of a lower signal-to-noise cluster of $M_{vir} =
9\times10^{13}h^{-1}M_\odot$ at the same redshift. As expected, in both cases we see a concentration of detections around the
centre of the image, and a roughly uniform background with the
indication of some edge effects on the borders of the image.

\begin{figure}
\includegraphics[width=0.24\textwidth]{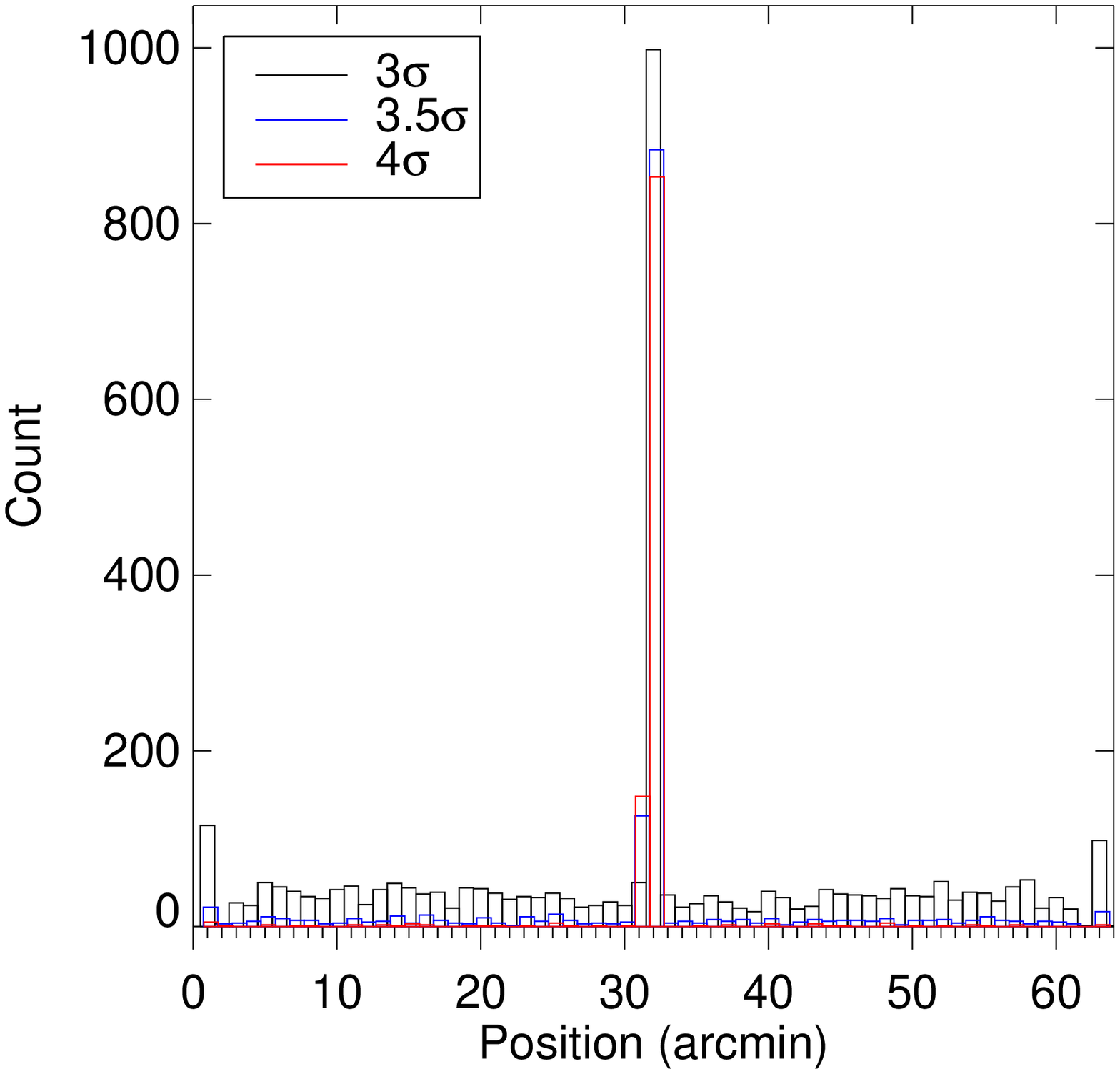}\includegraphics[width=0.24\textwidth]{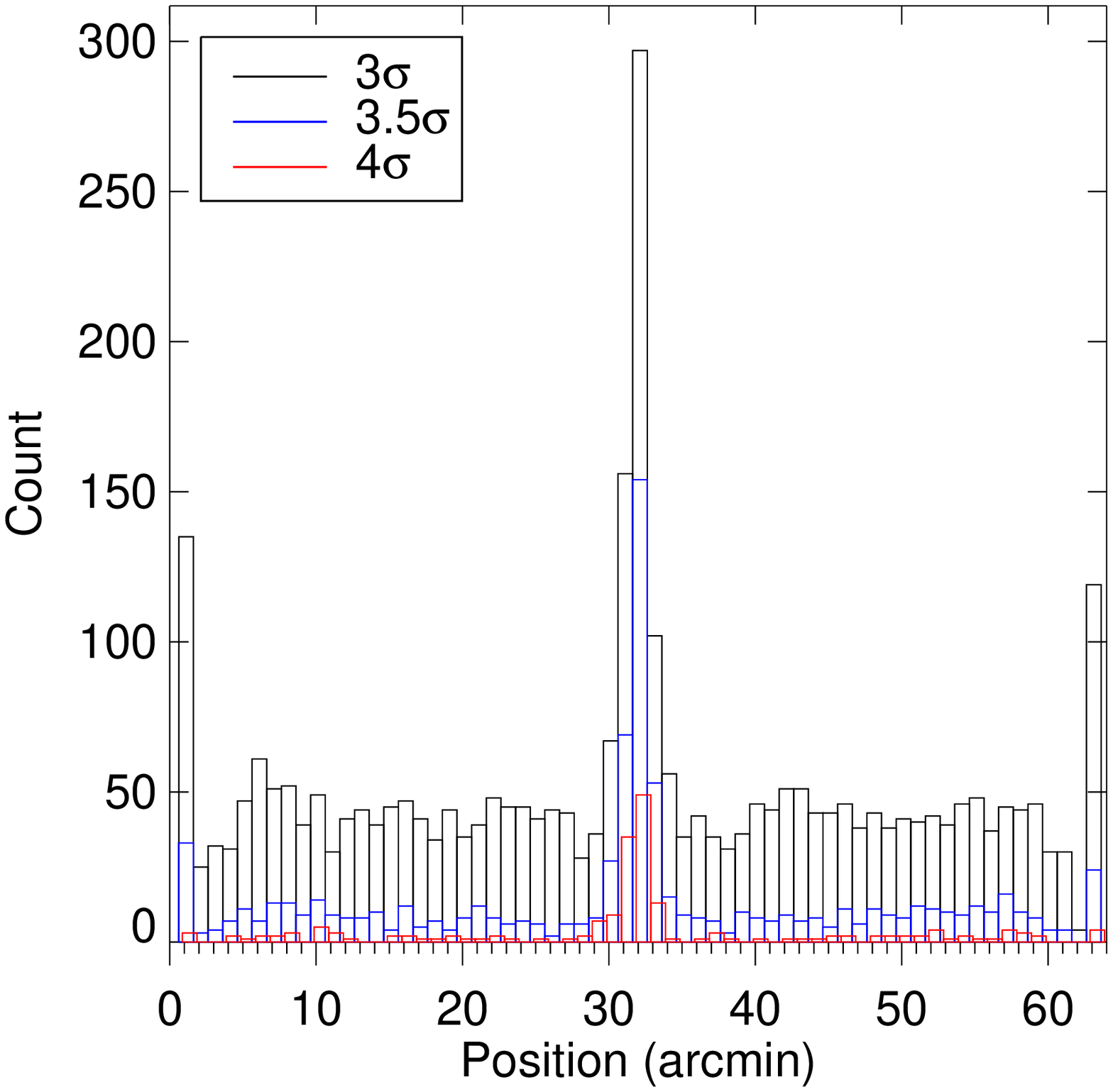}
\caption{This figure shows the distribution of the $x$ positions ($y$
  positions are assumed to be symmetrical) of peaks detected in 1000
  MRLens 2D reconstructions of a field containing a cluster of virial
  mass $7\times10^{14}h^{-1}M_\odot$ at a redshift of $z_{\rm
    cl}=0.35$ (\textit{left panel}) and the analogous plots for a field containing a smaller cluster of virial mass $9 \times 10^{13}h^{-1}M_\odot$ at the same redshift (\textit{right panel}). In both cases, the different coloured histograms are derived using
  different values of the denoising threshold. This choice clearly
  impacts both the false and true detection rates. \label{fg:xy2D}}
\end{figure}

Figure \ref{fg:xy2D} shows the comparable distribution of peaks from
MRLens 2D reconstructions of the same fields. Histograms are plotted
for the 3 different denoising thresholds as indicated in the
legend. In all cases, there is a concentration of detections
around the centre, as expected, and a background of false peaks. The
amplitude of both the central peak and the uniform background of false
detections decreases as the denoising threshold is raised. This is
expected: the higher the threshold used, the more we will suppress the
detection of peaks arising due to the noise. However, in doing this,
we also reduce somewhat the probability that we will detect the
central peak. 

When the edge effects at extremal values of $x$ and $y$ are excluded, the distributions shown in figures \ref{fg:xy3D} and \ref{fg:xy2D} can be fit by a Gaussian plus a constant offset representing the uniform background of false detections. We carried out such fits for the distributions of $x$ and $y$ generated from the 1000 reconstructions of each cluster field using GLIMPSE, and for MRLens for each of the three denoising thresholds tested. In all but the lowest signal-to-noise cases, where the fitting procedure failed, the mean $x$ and $y$ identified by the fitting procedure were clustered around $[\overline{x},\overline{y}] = [32,32]$, and the standard deviations $\sigma_x$ and $\sigma_y$ were consistently found to be in the range $0.3 - 1.4$, with a trend to larger values of the standard deviation for fields with a lower signal to noise. 

Given the difficulty of obtaining a reliable Gaussian fit for cluster fields where the rate of detection of the cluster is very low, we chose to define a cluster detection in both 2D and 3D
as any detection with $30 \le x \le 34$ and $30 \le y \le 34$ in pixel
units. Anything outside of this range is considered to be a false
detection. This choice may artificially reduce the true detection count in lower signal-to-noise fields relative to those with higher lensing signal-to-noise; however, in the majority of cases, we retain as true detections those peaks lying within $2-4$ standard deviations from the mean position at the centre of the field. Moreover, the width of the $x-y$ distributions for a given cluster field were found to be very similar regardless of the reconstruction method and denoising threshold. Therefore, the choice to restrict detections of the cluster to those falling within the range $30 \le x \le 34\ \& \ 30 \le y \le 34$ does not unfairly (dis)advantage any particular method.

We also note in the figures that there appears to be an
overdensity of detections near the edges of the reconstructions. In
order to minimise the contamination from these edge effects in the
analysis that follows, we exclude all false detections located within
4 pixels of the edge of the reconstruction in the transverse plane.

\begin{figure}
\includegraphics[width=0.25\textwidth]{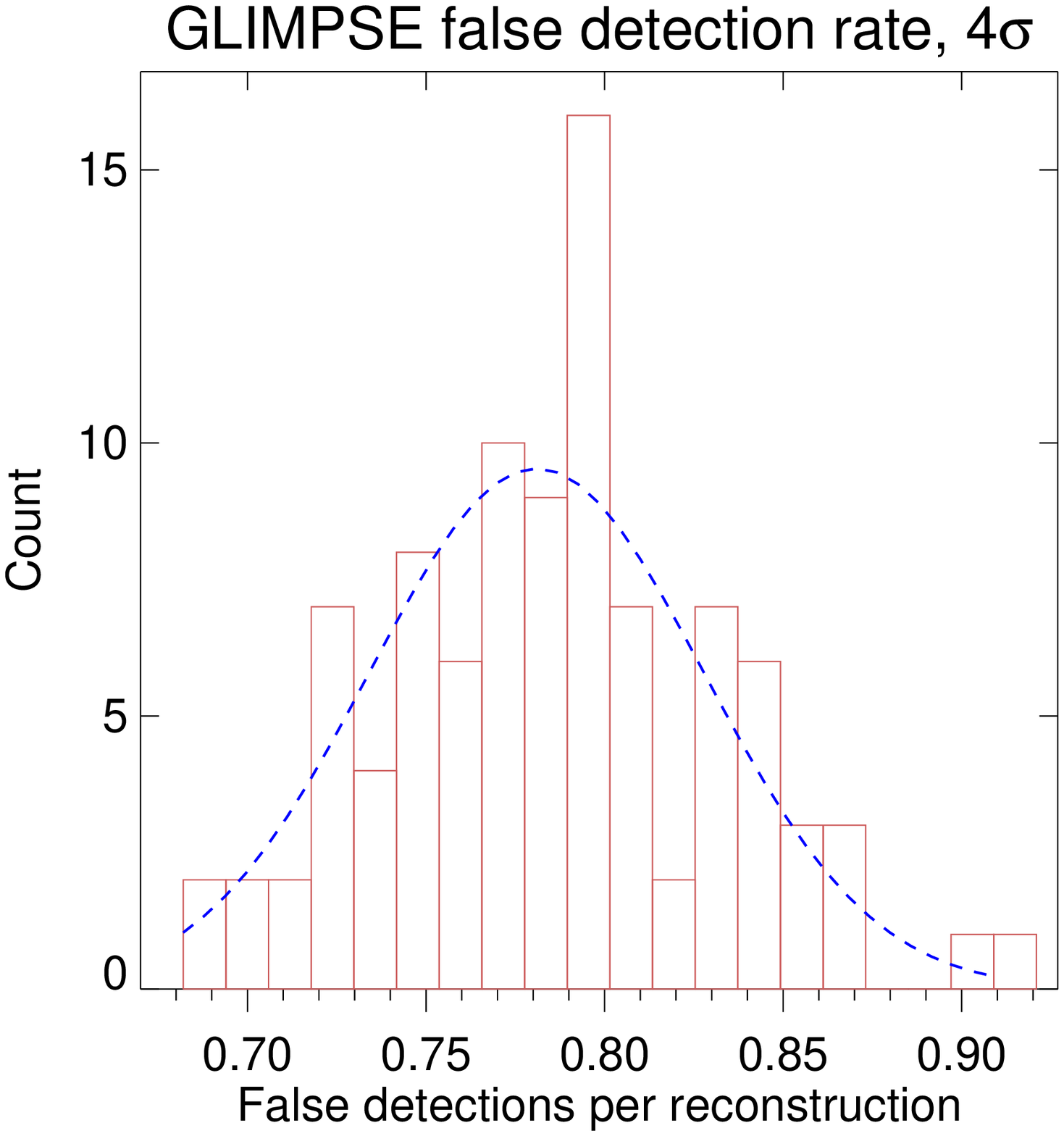}\includegraphics[width=0.25\textwidth]{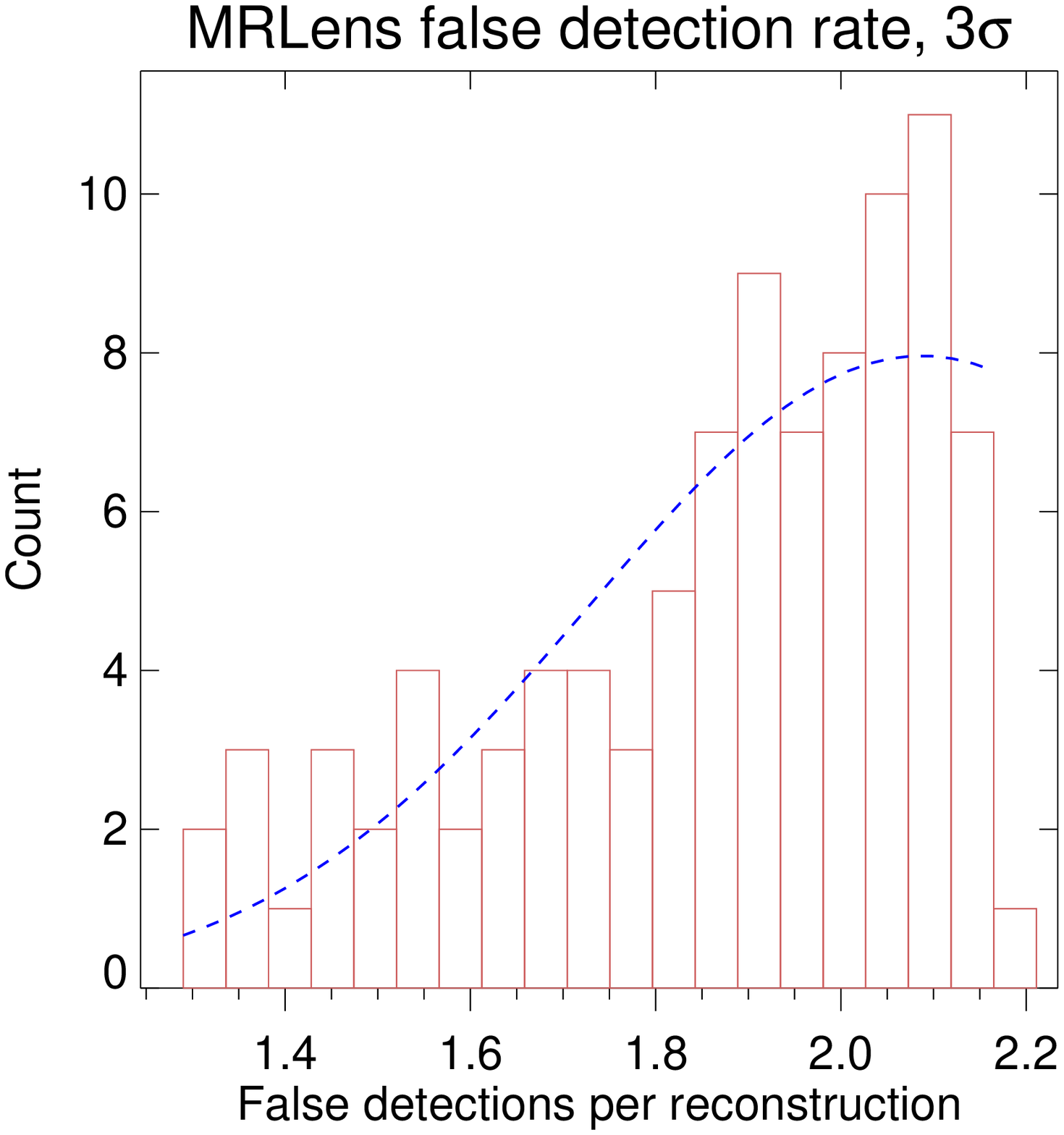}
\includegraphics[width=0.25\textwidth]{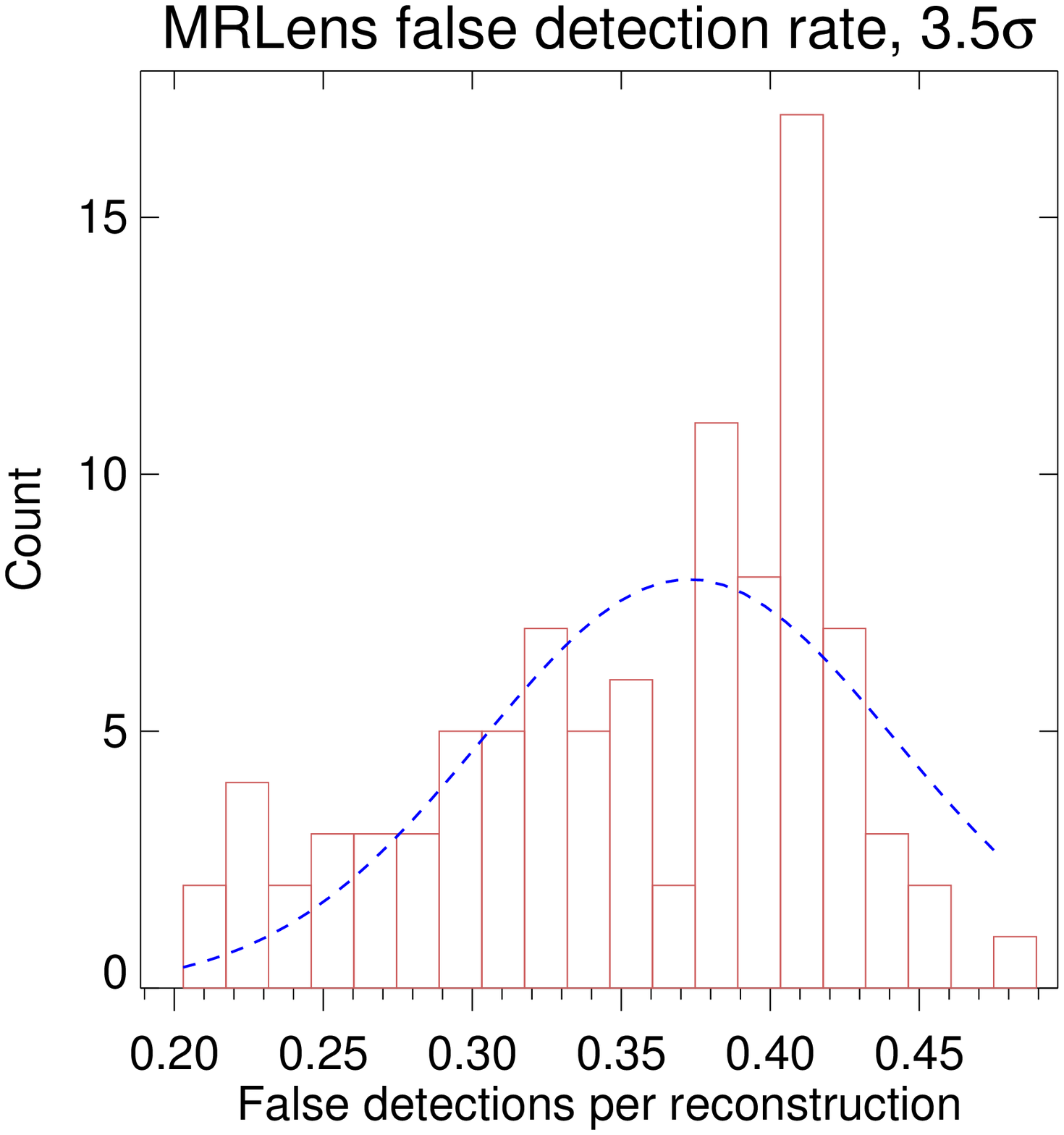}\includegraphics[width=0.25\textwidth]{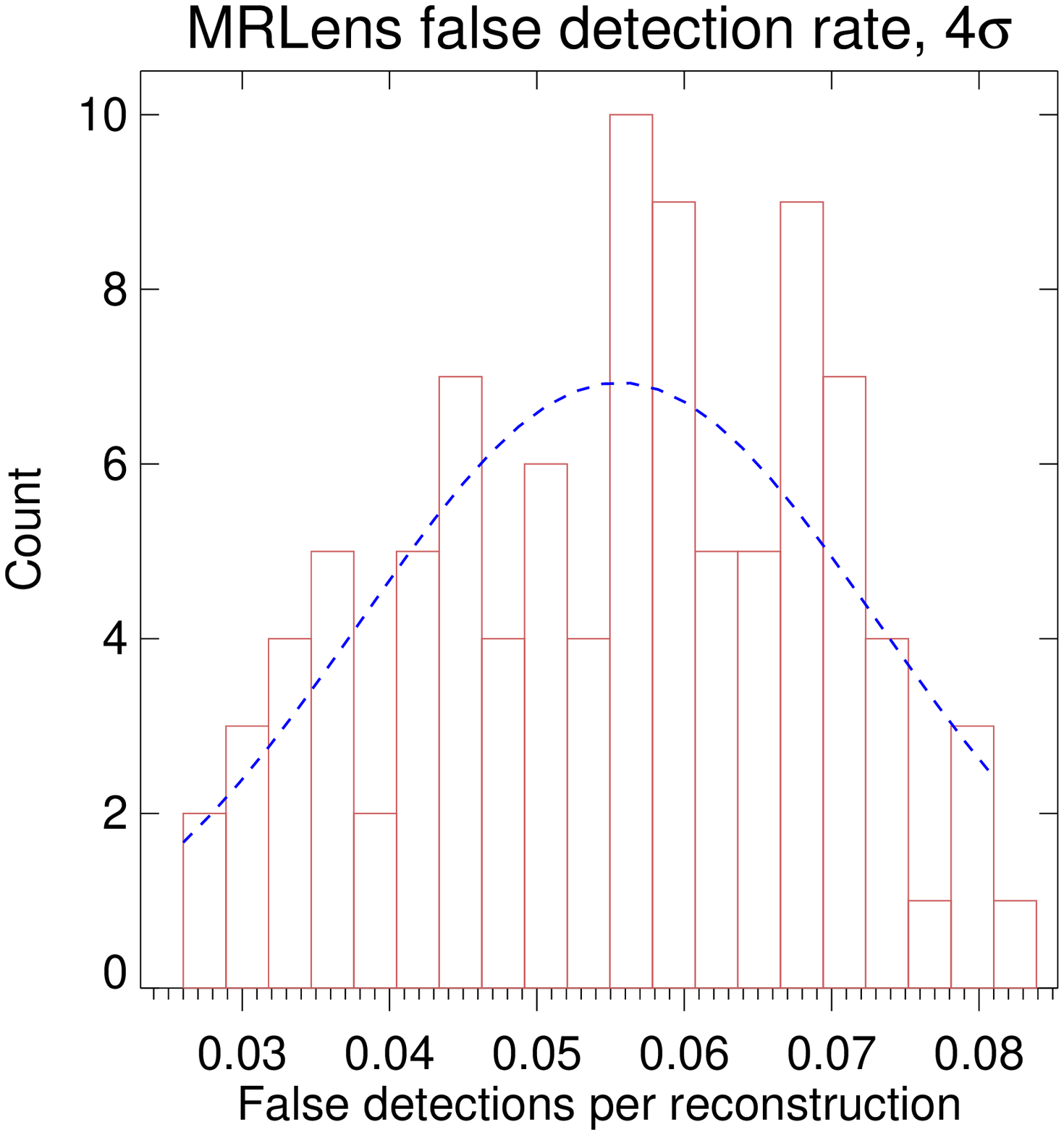}
\caption{This figure shows the distribution of the mean false
  detection rate per reconstruction for the 96 cluster fields
  described in the text, reconstructed using GLIMPSE with a $4\sigma$
  denoising threshold, and MRLens with denoising thresholds at
  $3\sigma$, $3.5\sigma$, and $4\sigma$. \label{fg:falsedets}}
\end{figure}

In figure \ref{fg:falsedets}, we show a histogram of the mean false
detection rate per reconstruction seen in each of our 96 cluster
fields using GLIMPSE, and using MRLens with denoising thresholds of
$3\sigma$, $3.5\sigma$ and $4\sigma$. Overplotted in the figures are
the best-fit Gaussian distributions. The fit parameters, the mean
$\mu_{\rm fit}$ and the standard deviation $\sigma_{\rm fit}$ are
listed in Table \ref{tab:falsedets} and compared with the median and
median absolute deviation (MAD) statistics computed from the 96
computed false detection rates.

\begin{table}
\begin{centering}
\begin{tabular}{l c c c c c }
\hline
Algorithm & Threshold & $\mu_{\rm fit}$ & $\sigma_{\rm fit}$ & Median & MAD \\ \hline
MRLens & 3$\sigma$ & 2.087 & 0.357 & 1.923 & 0.231 \\ 
MRLens & 3.5$\sigma$ & 0.373 & 0.069 & 0.377 & 0.062\\
MRLens & 4$\sigma$ & 0.056 & 0.018 & 0.056 & 0.018 \\
GLIMPSE & 4$\sigma$ & 0.781 & 0.047 & 0.786 & 0.044 \\ \hline
\end{tabular}
\caption{This table gives the parameters of the best-fit Gaussian to
  the distribution of the number of false detections per realisation
  in the 96 cluster fields analysed with MRLens at 3 different
  denoising thresholds and GLIMPSE at a denoising threshold of
  $4\sigma$. Also listed are the median and median absolute deviation
  (MAD) statistics computed from the 96 cluster field false detection
  rates. \label{tab:falsedets}}
\end{centering}
\end{table}

We can clearly see that the MRLens false detection rate in 2D is
substantially smaller (by a factor of $\sim 38$) than that seen in 3D
at the same denoising threshold of $4\sigma$, and only becomes
comparable to the $4\sigma$ GLIMPSE false detection rate when the
denoising threshold drops between $3-3.5\sigma$. As noted in section
\ref{subsec:noisepeaks}, given that there are more pixels in 3D than
in 2D, it is expected that we will have more false detections in 3D
than in 2D for the same threshold level, so this result is unsurprising.

On the other hand, it does indicate that when assessing the overall performance of the GLIMPSE and MRLens algorithms it is not sufficient to simply look at their cluster detection rates at the same denoising threshold. In both cases, there will be a trade-off between increasing the cluster detection rate by lowering the denoising threshold, and controlling the false detection rate with a stricter choice of threshold. However, in 2D the false detection rate is naturally lower than in 3D, and therefore a lower threshold can be chosen in 2D than in 3D for the same level of false peak contamination. 

For this reason, we compare the GLIMPSE results using a denoising threshold of $4\sigma$ with MRLens results at three different thresholds. Comparing the MRLens and GLIMPSE true detection rates when both use a $4\sigma$ threshold allows us to determine whether there is any intrinsic signal-to-noise advantage in studying lensing systems in 3D rather than 2D. Any information gained in 3D comes at the price of increased false detections, and so we lower the threshold in 2D so that the false detection rates become comparable between GLIMPSE and MRLens, to determine whether the information gain in 3D is sufficient to outweigh the increased false detections compared with what is achievable in 2D.

\subsection{Control of false detections}

It is clear that spurious peaks arising from the noise may mimic real peaks in a reconstruction, and will therefore be a significant contaminant in any study that aims to constrain cosmological parameters using peak counts, or which aims to estimate the mass function. It is therefore important to consider strategies that might be implemented to remove these false detections or limit their impact on any resulting cosmological inference. 

Figures \ref{fg:xy3D} and \ref{fg:xy2D} indicate
that the distribution of false peaks is uniform across the field under
randomisations to the underlying noise. While this is true in our idealised simulations, in which we have assumed a uniform distribution of background sources, real data often contains a complex gap structure and boundaries, which may give rise to a less uniform distribution of false peaks. We nonetheless expect that peaks arising due to the noise will not appear consistently in the same places under different, randomised realisations of the noise.

In practice, when applying either GLIMPSE or MRLens to real data, one would need to carry out Monte$-$Carlo or bootstrap resampling of the data in order to gain a full understanding of the noise properties of the reconstructed maps. Moreover, as demonstrated in \cite{LLS14a}, a reliable estimate of the mass and redshift of a detected cluster, and their associated errors, can only be obtained by considering an ensemble of reconstructions carried out under randomisations to the noise in the data.

\begin{figure}
\centering
\includegraphics[width=0.14\textwidth]{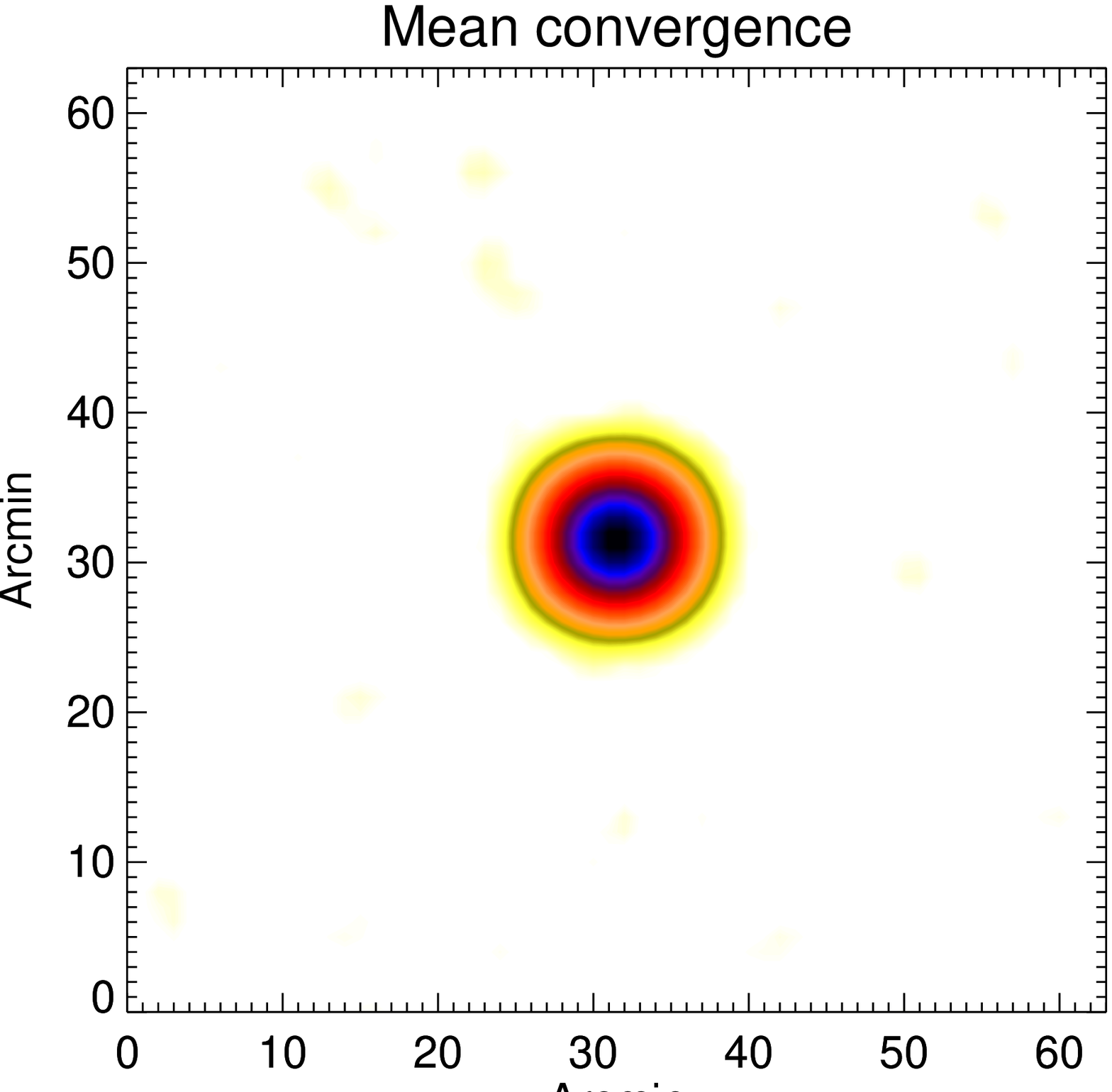}\hskip10ex\includegraphics[width=0.2\textwidth]{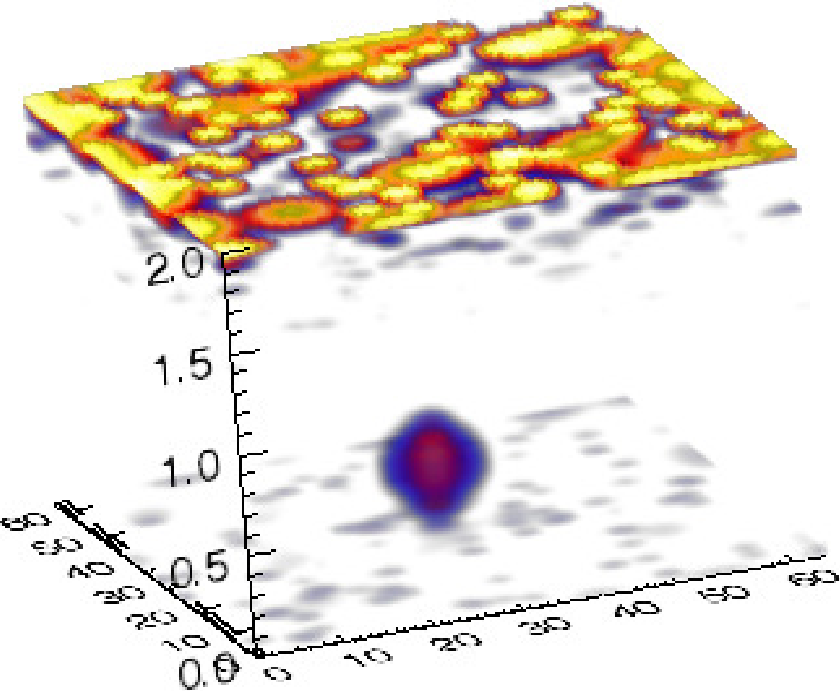}
\caption{The mean of 1000 reconstructions in 2D (3$\sigma$ threshold;
  \textit{left panel}) and 3D (4$\sigma$ threshold; \textit{right
    panel}) of field 53, containing a cluster of mass $M_{vir} =
  7\times10^{14}h^{-1}M_\odot$ and redshift $z_{\rm cl}=0.35$. Both
  plots use a logarithmic colour scheme, to accentuate any low-level
  false detections in the mean reconstruction. The detection rate for
  this cluster in both 2D and 3D at the chosen denoising thresholds is
  $100\%$.\label{fg:meandens53}}
\end{figure}

Given that false detections are randomly distributed, the probability of detecting a noise peak repeatedly in the same position under a different realisation of the noise is very low. In other words, we expect the 'detection rate' associated with a given false detection at a position $[x,y,z]$ to be very low. One way to visualise this is to consider a mean reconstruction, obtained by summing the reconstructions obtained from the all 1000 realisations of the data. 

A real cluster should be detected consistently at the same location and with a similar peak amplitude in (many of the) reconstructions using different realisations of the noise, so this feature should be prominent in such a mean map, while the false detections, which appear in random positions, will be suppressed by averaging many reconstructions. 

\begin{figure}
\centering
\includegraphics[width=0.14\textwidth]{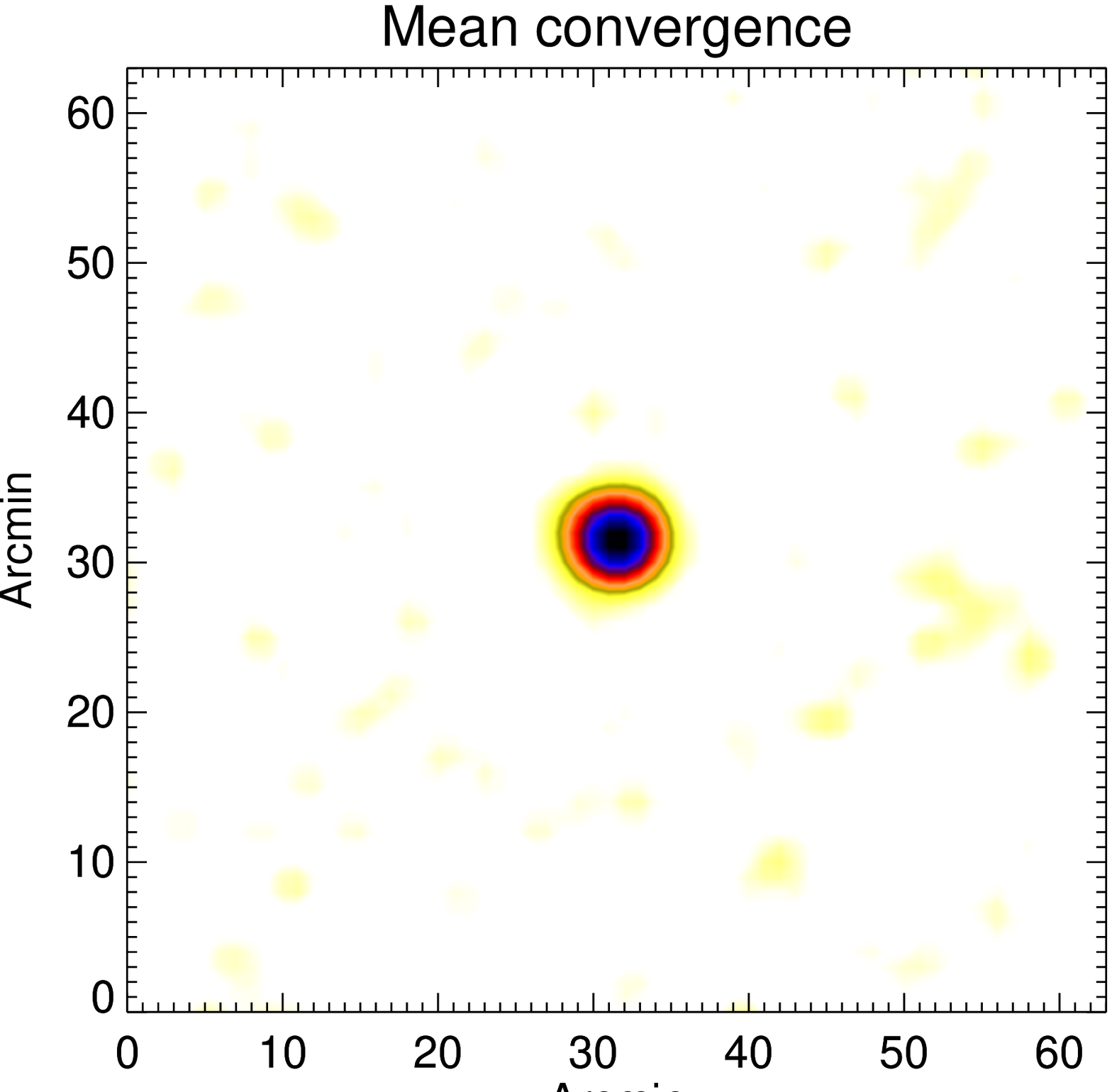}\hskip10ex\includegraphics[width=0.2\textwidth]{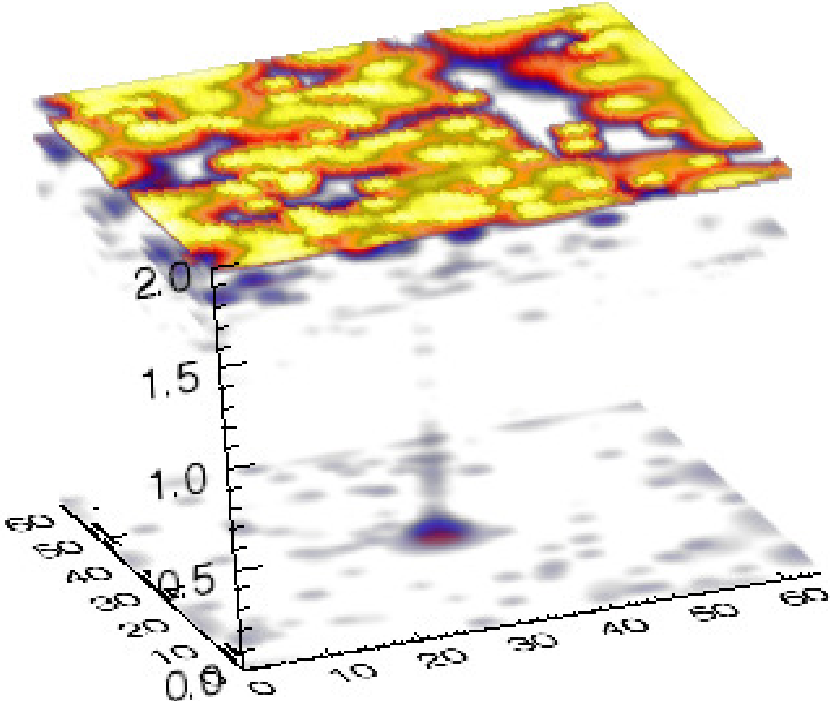}
\caption{The mean of 1000 reconstructions in 2D (3.5$\sigma$ threshold;
  \textit{left panel}) and 3D (4$\sigma$ threshold; \textit{right
    panel}) of field 46, containing a cluster of mass $M_{vir} =
  9\times10^{13}h^{-1}M_\odot$ and redshift $z_{\rm cl}=0.35$. Again,
  we use a logarithmic colour scheme. The detection rate for this
  cluster in both 2D and 3D at the chosen denoising thresholds is
  $\sim 50\%$. \label{fg:meandens46}}
\end{figure}

Figure \ref{fg:meandens53} shows the mean 2D (3.5$\sigma$ threshold)
and 3D (4$\sigma$ threshold) reconstructions of field 53. To highlight
any low-level false detections, we plot the densities on a logarithmic
colour scale. While this particular cluster has a detection rate of
$\sim 100\%$ in both 2D and 3D, the mean density maps are typical for
clusters with a much lower detection rate (e.g. figure
\ref{fg:meandens46}, where we show the results for field 46,
containing a cluster of mass $M_{vir} = 9\times10^{13}h^{-1}M_\odot$,
redshift $z_{\rm cl}=0.35$, and detection rate $\sim 25\%$). In 2D, it
is trivial to identify the location of the true density peak in the
mean reconstruction: no other significant peaks are visible in the
mean map.

In 3D, we notice several interesting features. Firstly, the cluster is
clearly visible in the mean reconstruction, but appears to be smeared
out in redshift. This is because the broad lensing efficiency kernel
$\mathbf{Q}$ gives rise to a large uncertainty in the redshift at
which the cluster is detected. This is evidenced by the large error
bars seen in the redshift estimates for these clusters presented in
\cite{LLS14a}. The cluster is clearly defined, though, and no
significant false detections are seen at redshifts below $z\lesssim
1.5$.

We do see the significant presence of high-amplitude false detections
at the high-redshift boundary of the mean reconstruction,
however. Such false detections were seen in \cite{LDS12} and
\cite{LLS14a}, and arise due to overfitting of the shear data in the
highest redshift bins.

\begin{figure}
\includegraphics[width=0.45\textwidth]{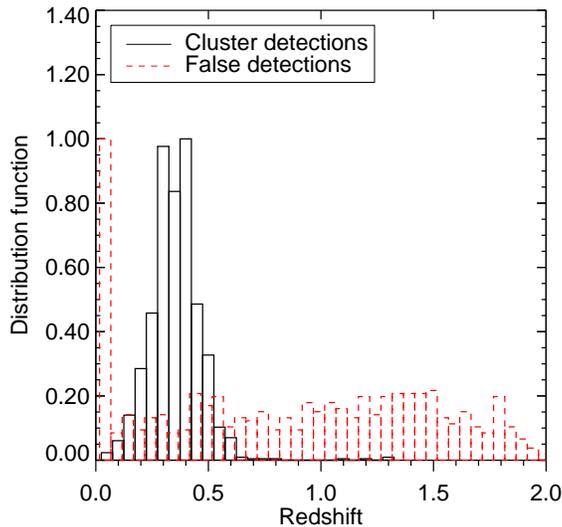}
\caption{The redshift distribution of cluster detections and false
  peaks in reconstructions of 1000 noisy realisations of shear data
  for field 53, containing a cluster of mass $M_{vir}
  =7\times10^{14}h^{-1}M_\odot$, redshift $z_{\rm
    cl}=0.35$. \label{fg:zdist_dets}}
\end{figure}

In figure \ref{fg:zdist_dets}, we plot the redshift distribution of
the central cluster detections and the false detections for field
53. While the redshift distribution of the cluster detections shows an
approximately Gaussian distribution centred on the true input redshift
(in this case, $z_{\rm cl} = 0.35$), the redshift distribution of the
false detections shows a significant overdensity at very low and very
high redshift, and a flat distribution in between. This distribution
is characteristic of all the cluster fields studied.  Given that we
physically do not expect to see such a high density peak at such high
redshifts, truncating the 3D reconstructions at $z=1.5$ will remove
these high redshift false detections, which account for $\sim 30\%$ of
all false detections seen in 3D, whilst having a minimal impact
($<2\%$ on the cluster detection rate).

\subsection{Cluster detection rate: 2D vs 3D}
\label{subsec:results}

\begin{figure*}
\begin{center}
\includegraphics[width=0.33\textwidth]{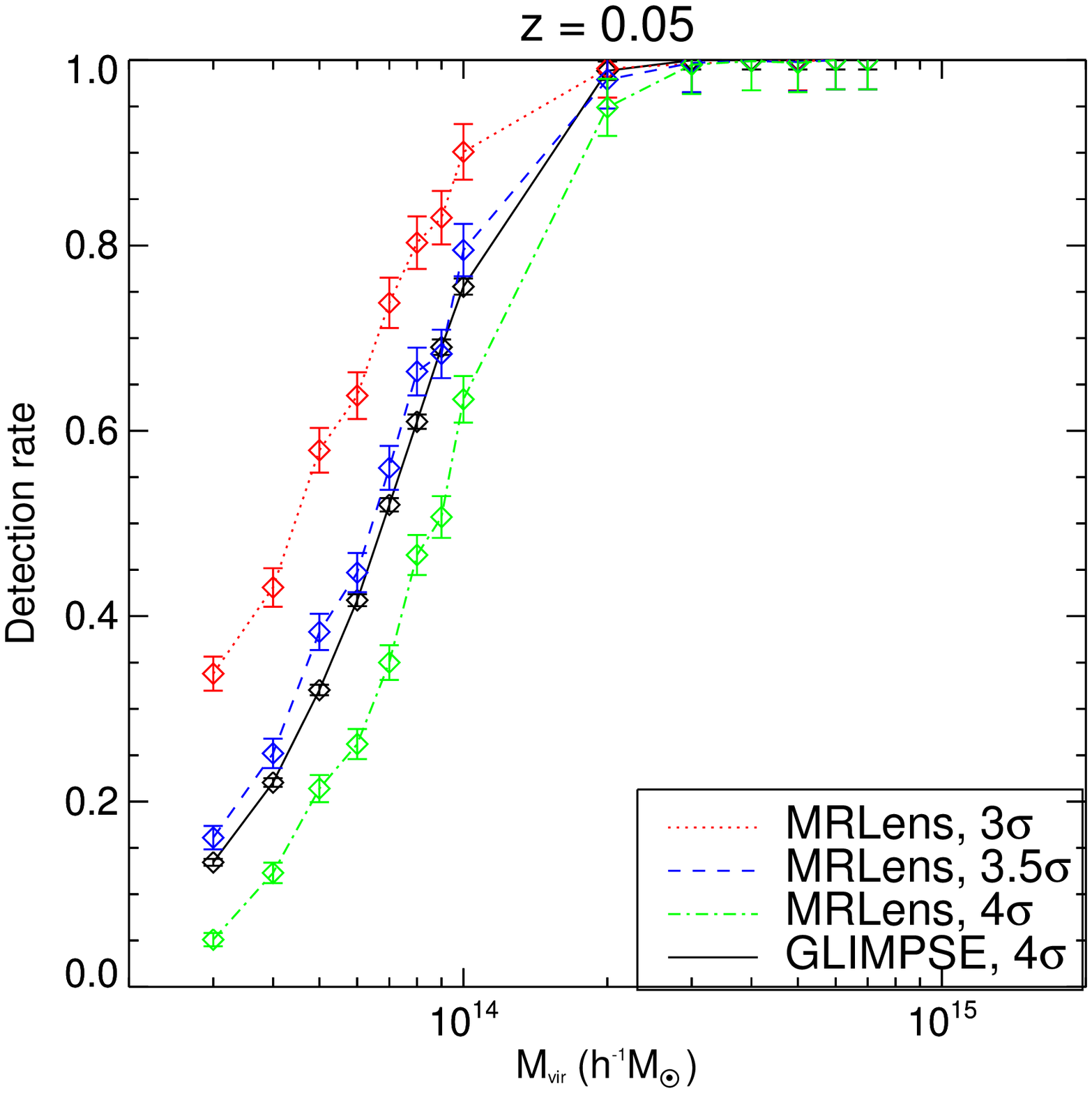}\includegraphics[width=0.33\textwidth]{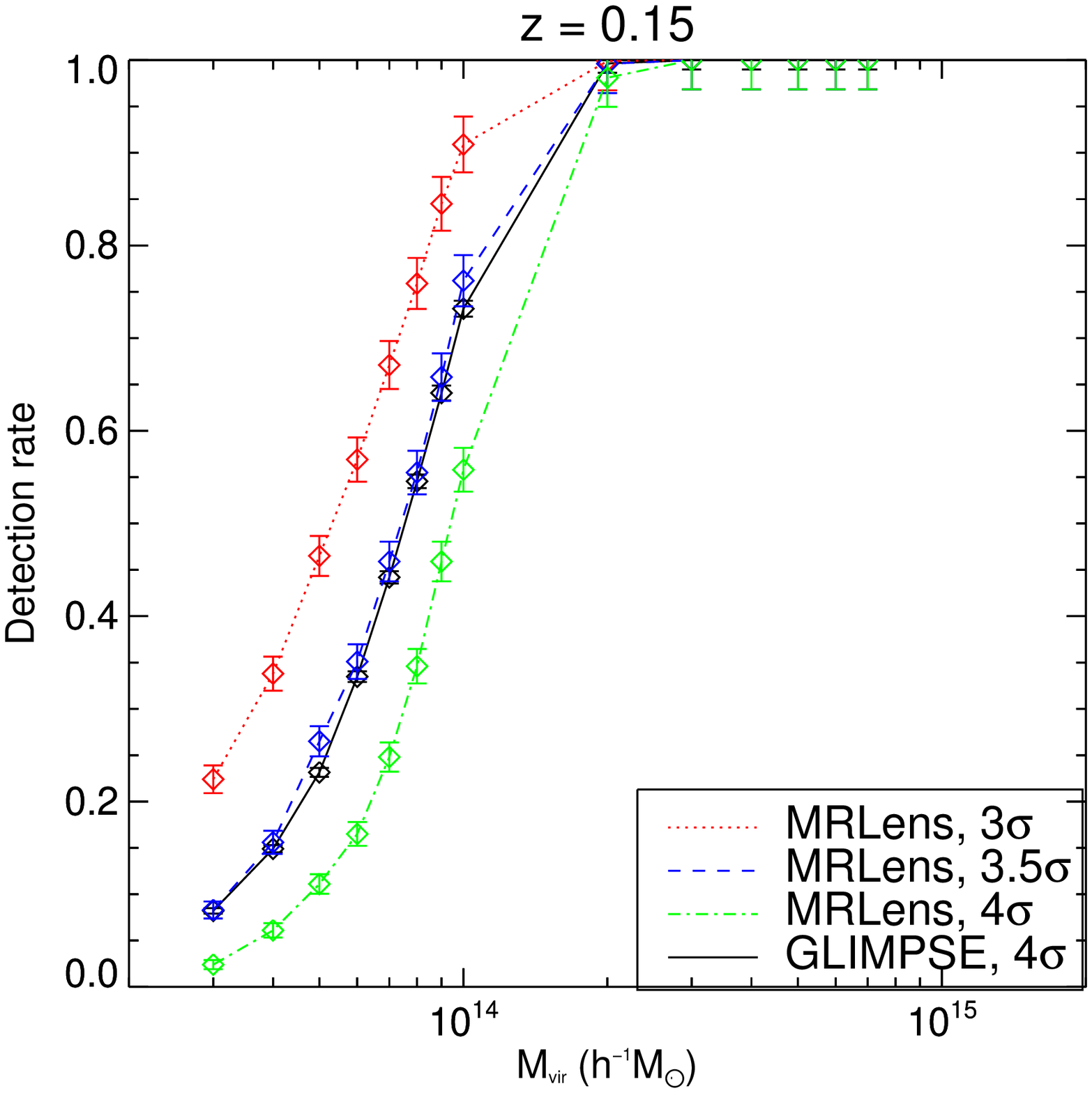}\includegraphics[width=0.33\textwidth]{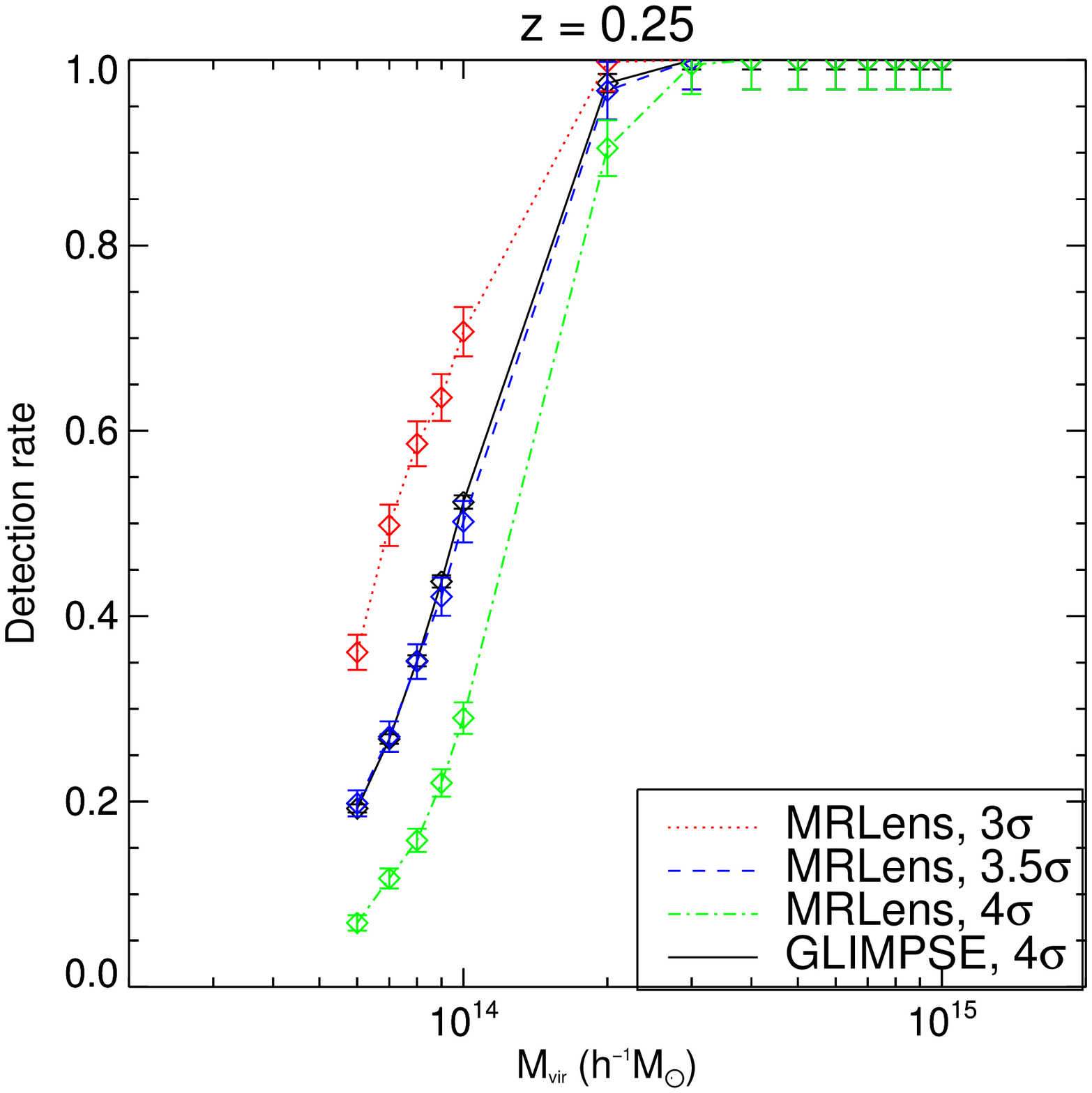}
\includegraphics[width=0.33\textwidth]{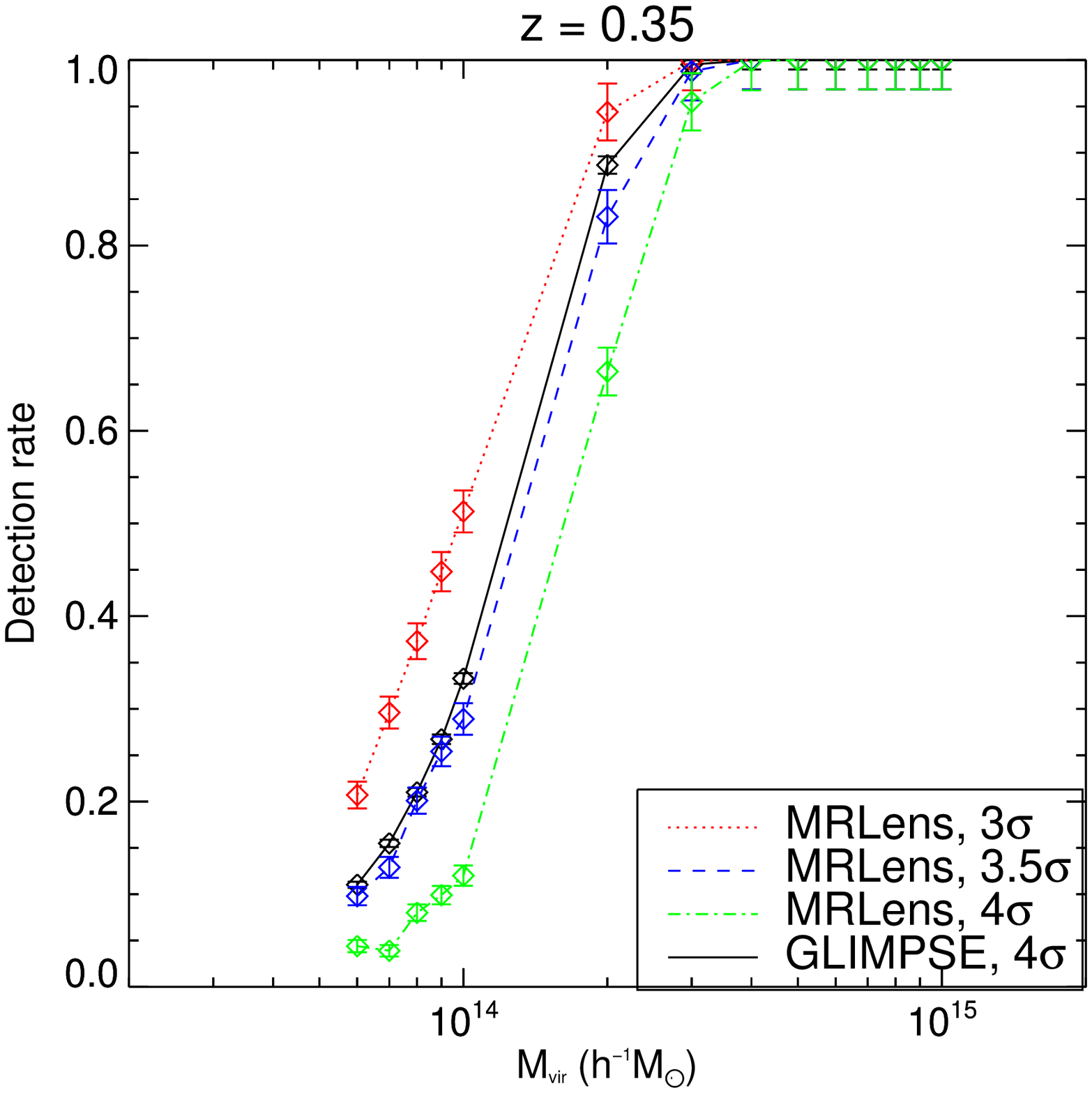}\includegraphics[width=0.33\textwidth]{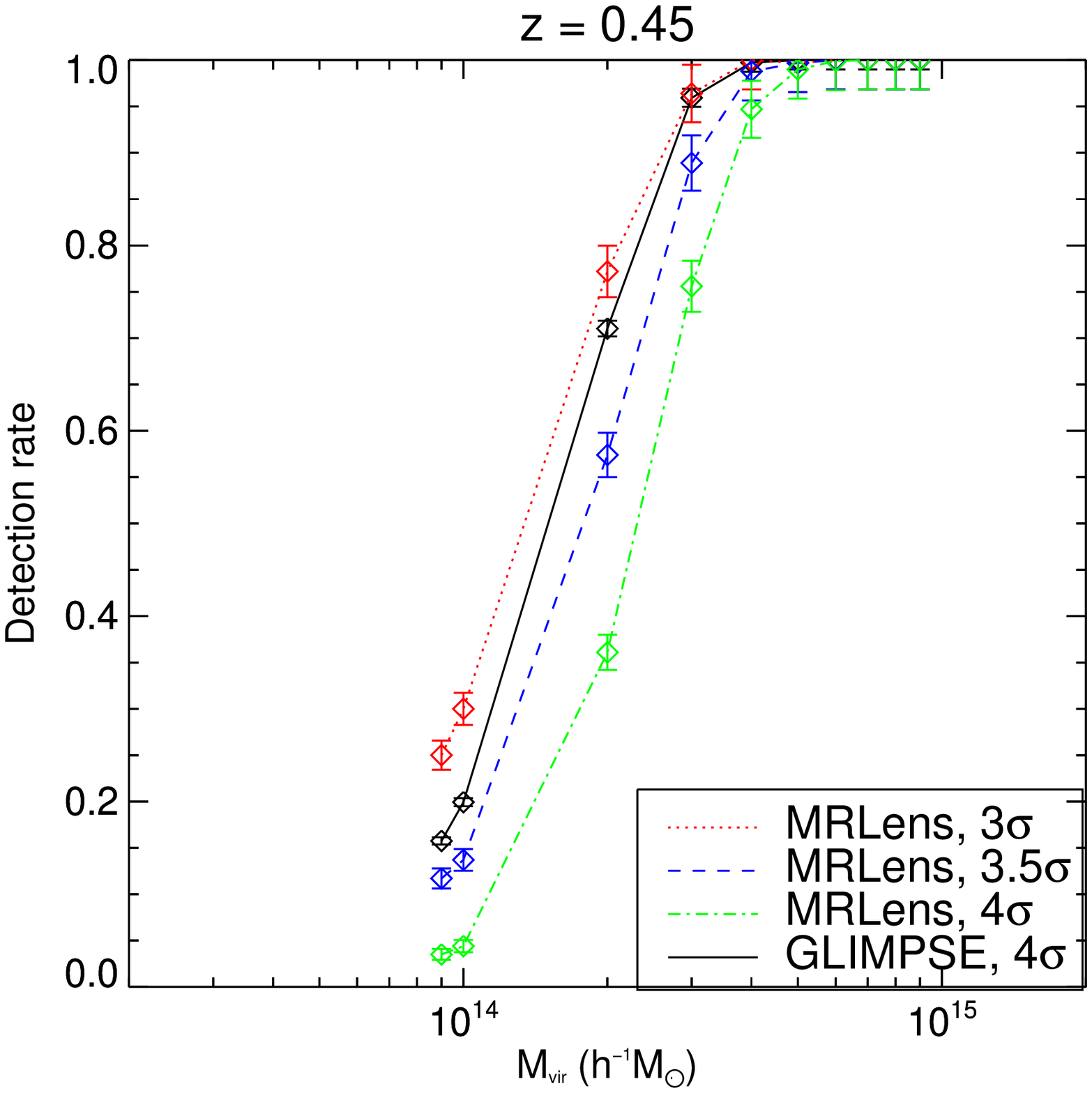}\includegraphics[width=0.33\textwidth]{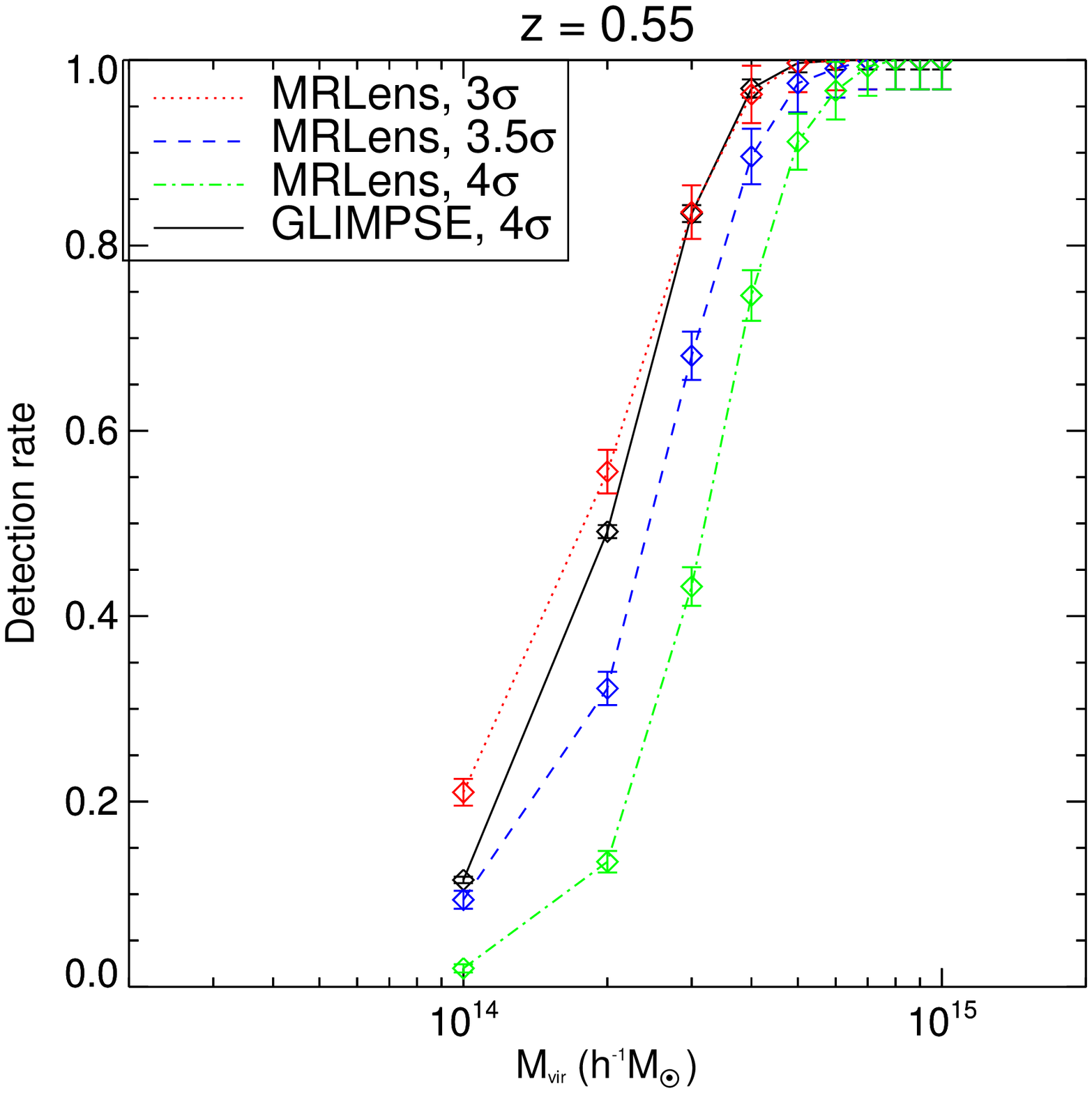}
\includegraphics[width=0.33\textwidth]{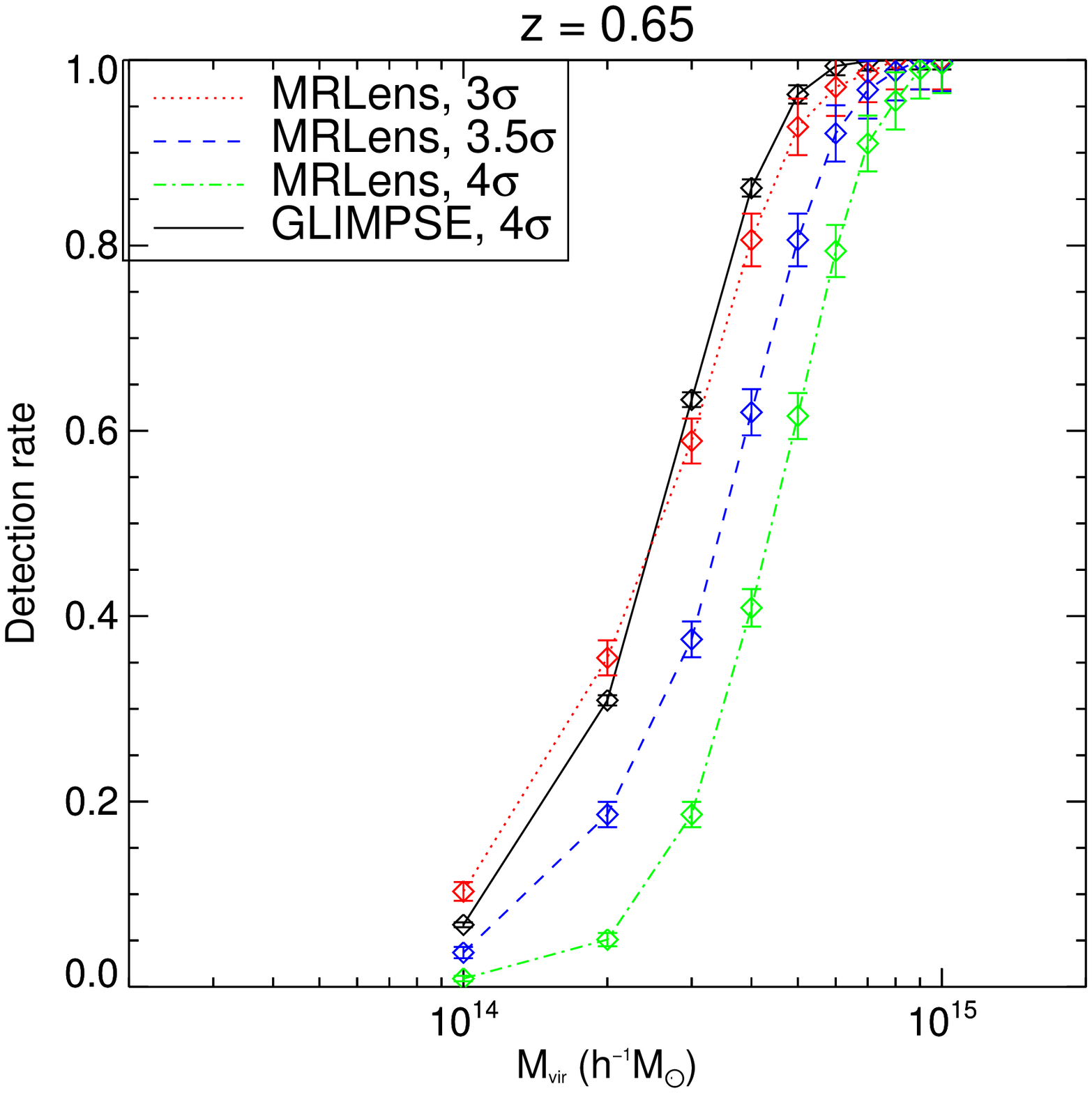}\includegraphics[width=0.33\textwidth]{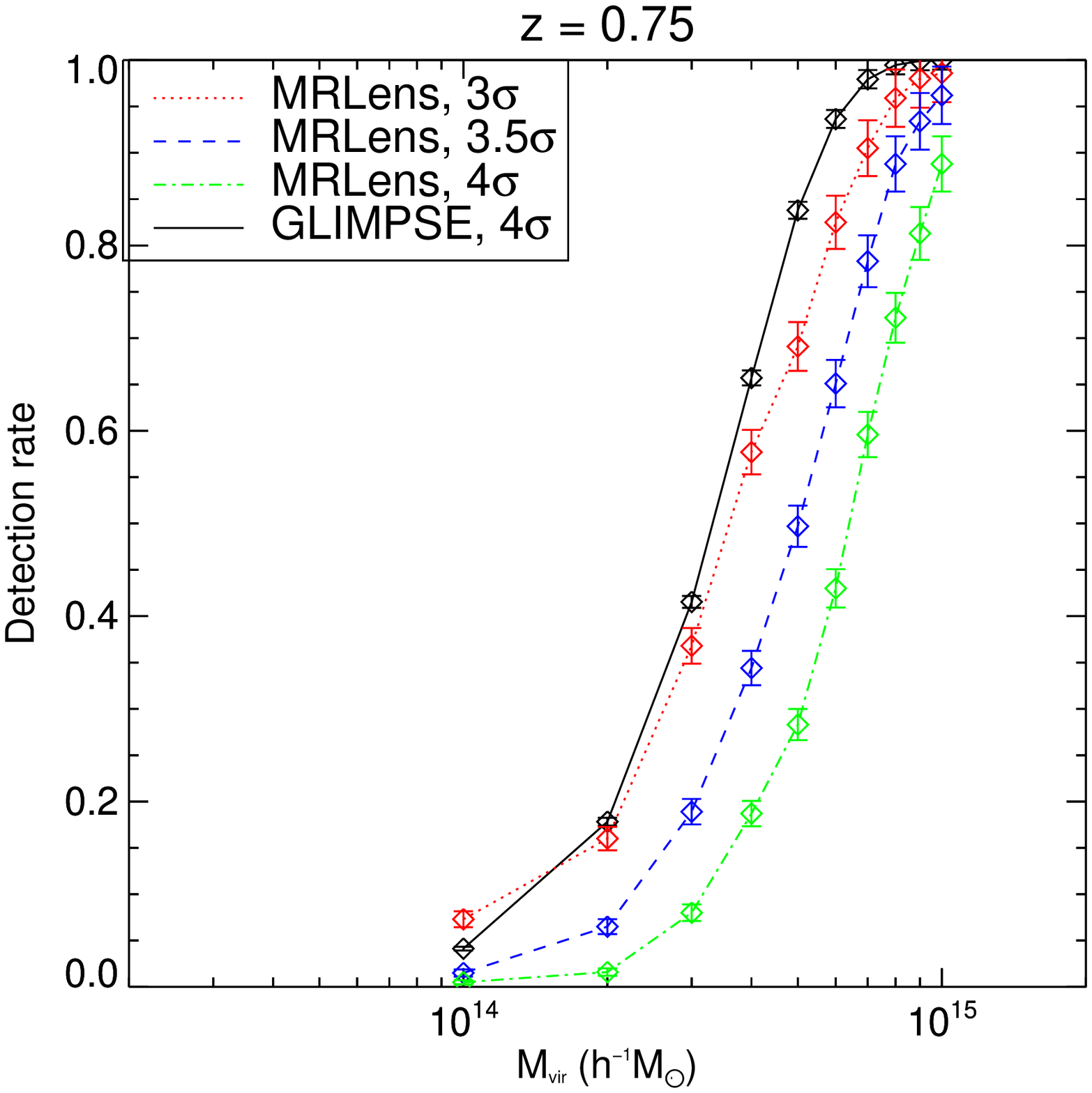}
\caption{The fraction of noise realisations in which the central
  cluster was detected as a function of mass and redshift. 3D mapping
  clearly outperforms 2D mapping for detection of clusters for all but
  the highest mass clusters at which both methods asymptote to a
  detection rate of $100\%$.\label{fg:selfuncs}}
\end{center}
\end{figure*}

With false detections now well understood, we turn our attention to
consider the detection rate expected for clusters as a function of
mass and redshift for both MRLens and GLIMPSE. Figure
\ref{fg:selfuncs} shows the fraction of noise realisations in which a
given cluster was detected as a function of mass for each of the 8
different redshifts at which we simulated cluster haloes. This figure
effectively shows the selection function for each method in the
configurations specified above as a function of mass and
redshift. Error bars were computed assuming Poisson noise only.

The selection functions in 2D and 3D all show the same general trends:
an asymptote towards zero detections at low mass and an asymptote at
high mass towards 100\% detections. There is also a consistent trend towards lower detection rates at high redshift than low redshift for fixed cluster mass. However, these dependencies differ between 2D and 3D:
\linebreak
\linebreak
\noindent\textbf{GLIMPSE (4$\sigma$) vs MRLens (4$\sigma$):\\} 
\indent This comparison offers insight into whether there is a natural signal-to-noise advantage in studying weak lensing clusters in 3D rather than 2D. In 26 fields, both GLIMPSE and MRLens have detection rates of $100\%$ (the GLIMPSE detection rate is $100\%$ in 39 of the cluster fields). In every other field, GLIMPSE outperforms MRLens in terms of detection rate. The largest improvement is seen for field 88, a $2\times10^{14}h^{-1}M_\odot$ cluster at a redshift of $0.75$, where the GLIMPSE detection rate of 0.178 is a factor of 11.1$\times$ larger than that with MRLens (0.016). The GLIMPSE detection rate is $2\times$ higher than that of MRLens in 27 fields and $5\times$ in 6 fields. There therefore appears to be a natural advantage to using 3D reconstructions for cluster detection in weak lensing surveys. 

However, as noted in section \ref{subsec:algoconfig}, comparing the
GLIMPSE and MRLens detection rates at the same detection threshold is
not an entirely fair comparison, as the GLIMPSE reconstructions show a
$\sim 38\times$ higher rate of false detections than the MRLens 2D
reconstructions at a $4\sigma$ threshold. A more fair comparison can
be obtained by reducing the MRLens denoising threshold to a level such
that the reconstructions show a similar rate of false detections. This
level is reached at a threshold somewhere between $3-3.5\sigma$.
\linebreak
\linebreak
\textbf{GLIMPSE (4$\sigma$) vs MRLens (3.5$\sigma$):}\\ 
\indent The GLIMPSE false detection rate at $4\sigma$ is significantly closer (within a factor of $\sim 2$) to that of MRLens at $3.5\sigma$, so a comparison between these two sets of reconstructions is of particular interest. The detection rates in this case, particularly at low redshift, are quite similar. In 31 cluster fields, both MRLens and GLIMPSE attain a $100\%$ detection rate. In 17 cluster fields, the MRLens detection rate is higher than the GLIMPSE detection rate, but the difference is typically on the order of $5-10\%$, with the largest difference, $\sim 20\%$, seen in field 1 where the GLIMPSE detection rate is 0.134 and the MRLens detection rate is 0.161. As the cluster is moved to higher redshift, the detection rates in 3D become consistently higher than in 2D: in 48 fields, the GLIMPSE detection rate is higher than that of MRLens, and in 3 of those fields the improvement is greater than a factor of 2. Field 88 is again one of the fields in which the largest improvement is seen. In this case the GLIMPSE detection rate is $\sim 2.7\times$ larger than that of MRLens.
\linebreak
\linebreak
\textbf{GLIMPSE (4$\sigma$) vs MRLens (3$\sigma$):}\\ 
\indent Given that the false detection rate of GLIMPSE at $4\sigma$ remains significantly higher than that of MRLens at $3.5\sigma$, for completeness we again lower the MRLens denoising threshold to $3\sigma$. Here the MRLens false detection rate is $\sim 2.5\times$ larger than that of GLIMPSE. With such a low denoising threshold in 2D, we see a corresponding boost in detection rates, as expected. In this case, we find that both GLIMPSE and MRLens attain a $100\%$ detection rate in 34 cluster fields. In 44 of the cluster fields, the MRLens detection rate is higher than that of GLIMPSE. These are primarily at lower redshift, and the detection rates remain within a factor of $2$ for all but 4 of the fields. The largest difference is seen in field 15, where the MRLens detection rate (0.224) is $2.7\times$ larger than that of GLIMPSE (0.082). At higher redshift, we find that the GLIMPSE detection rate remains higher than that of MRLens in 20 of the cluster fields, but the improvement here is moderate (the ratio between the GLIMPSE and MRLens detection rates is typically of order $\sim 1.2$).

\section{Discussion}
\label{sec:discussion}

The results above clearly demonstrate that mass mapping is a very
useful tool for detecting clusters of galaxies in weak lensing
surveys. At low redshift, GLIMPSE and MRLens provide
reconstructions of comparable quality, when one considers both the
cluster detection rate and the false detection rate seen in these
reconstructions. However, GLIMPSE mapping offers a distinct advantage
in the detection of clusters at high redshift, showing a significantly
higher sensitivity to such clusters than seen in 2D with MRLens.

The improvement seen in 3D is likely due to a
  combination of effects. In projecting the data down into 2D, we lose
  some of the information content contained in the data. Additionally, particularly in the
  case of high redshift clusters, the lensing signal becomes damped
  when averaging galaxy shapes over the full range of redshifts probed
  by the survey: in our experiment, the average ellipticity in each
  angular pixel $[x,y]$ is given by
\begin{equation}
\epsilon(x,y) = \frac{\int_0^2 p(z)\gamma(x,y,z) dz}{\int_0^2 p(z) dz}+\mathcal{N}(0,\sigma_{\varepsilon}/\sqrt{n})\ ,
\label{eq:meanell}
\end{equation}
where $\gamma(x,y,z)$ is in this case the analytically-derived shear at angular position $[x,y]$ and redshift $z$, $p(z) \propto n(z)$ gives the probability distribution function for the galaxies in the survey and $\mathcal{N}$ is a random noise contribution drawn from a Gaussian of mean 0 and standard deviation $\sigma_\varepsilon/\sqrt{n}$, where $n$ is the number of galaxies per pixel. In practice, in a real galaxy survey, equation \eqref{eq:meanell} will be replaced by a discrete sum over galaxy ellipticities. 

Clearly, the lensing signal $\gamma(x,y,z) \equiv 0\ \forall z<z_{\rm cl}$, where $z_{\rm cl}$ is the redshift of the cluster. If the cluster redshift is known \textit{a priori}, one can retain only those galaxies behind the cluster; however in the absence of this knowledge, computing a 2D shear map by averaging over all the galaxies in a survey will result in a damping of the lensing signal due to the inclusion of galaxies that are not lensed, and this effect will be more pronounced for higher redshift clusters.

The GLIMPSE 3D reconstruction method is significantly more
sensitive than the state-of-the-art 2D reconstruction algorithm
MRLens, particularly for clusters at low mass and/or at high redshift.
At low redshift, the quality of GLIMPSE and MRLens reconstructions are comparable. The false detection rates are similar between GLIMPSE and MRLens at $3.5\sigma$, and the detection rates align similarly: GLIMPSE exhibits a detection fraction above $\gtrsim
50\%$ for clusters of mass $M_{vir}\gtrsim 10^{14}h^{-1}M_\odot$ at
low redshift ($z\sim0.25$); MRLens provides a comparable detection rate at a $3.5\sigma$ at this redshift with the detection rates at $3\sigma$ and $4\sigma$ being $70\%$ and $30\%$, respectively. At high redshift, there is a clear improvement in 3D reconstructions. For example, GLIMPSE detects $63\%$ of clusters of mass $3\times10^{14}h^{-1}M_\odot$, while the MRLens detection rates are $59\%\ (3\sigma),\ 37.5\%\ (3.5\sigma)\mbox{ and } 19\%\ (4\sigma)$. At high redshift, where the weak lensing signal-to-noise is low, it is therefore significantly advantageous to employ GLIMPSE for the detection of clusters in weak lensing maps.

We have demonstrated that MRLens and GLIMPSE can be tuned to yield
mass reconstructions with similar levels of purity; in other words,
reconstructions in which the contamination due to peaks arising from
the noise is at a similar level in both 2D and 3D reconstruction
methods. Given the findings of \cite{pls12} that peak counts
undertaken on MRLens reconstructions provide a powerful method to
probe non-Gaussianity and to discriminate between degenerate
cosmological models, it is reasonable to expect that 3D
reconstructions with GLIMPSE will offer an even more powerful
constraint on non-Gaussianity, given the higher levels sensitivity to
high-redshift clusters seen in GLIMPSE reconstructions.

Furthermore, as described in \cite{LLS14a}, GLIMPSE reconstructions
provide an unbiased estimator of both the masses and redshifts of the
clusters detected. We expect that the selection function for optical
halo finders will differ somewhat from that of GLIMPSE, but a detailed
comparison would require a sophisticated simulation involving N-body
cosmological simulations, raytracing, and a realistic semi-analytic
modelling of galaxy formation. This comparison will be the subject of
future work, and has implications for the study of galaxy clusters and
their statistics in large upcoming surveys such as Euclid.

A final cosmological application of GLIMPSE would be to measure the
evolution of the high-mass end of the mass function. To do this, we
need an accurate model of the GLIMPSE selection function as a function
of cluster mass, redshift, and concentration parameter, and also as a
function of the underlying cosmological model. Work is ongoing in this
area, and will be the subject of an upcoming publication. GLIMPSE is a
non-parametric reconstruction method, which means that we do not
compute a virial mass or redshift for the clusters we detect as part
of the algorithm, but rather compute masses by integrating the density
over a range of pixels in the reconstruction. In order to use GLIMPSE
outputs to compute the cluster mass function, we must first be able to
translate the GLIMPSE masses into virial masses, or to model the mass
function for the GLIMPSE observables as a function of virial mass,
redshift and cosmological model. This task is not trivial, but
certainly feasible.

It is important to note that the simulations presented here are highly idealised, and that real data will pose many additional challenges. The data themselves will contain complicated shear and photometric redshift systematics, and the effects of more complicated (and possibly correlated) errors, biases, and photometric redshift catastrophic failures, as well as the impact of intrinsic alignments, on the resulting mass maps will need to be investigated. 

Moreover, a more realistic density distribution containing clustered mass peaks will result in blending of structures in mass maps, and thereby impact the measured peak counts and/or cluster mass function. In a very preliminary work, \cite{LDS12} demonstrated that sparsity-based 3D mapping techniques may be able to disentangle different density peaks located along the same line of sight, and while extensive tests have not yet been carried out with GLIMPSE, the indications are that GLIMPSE improves greatly on the results of this earlier work. A full exploration of this is ongoing, but this may highlight yet another advantage of using 3D reconstructions, rather than two-dimensional projections in which the line-of-sight information is lost. 

In addition, both magnification and instrumental effects will give rise to a spatially-varying survey depth, and in combination with the removal of foreground objects such as stars, this will give rise to a very complex survey mask that must be accounted for. It is clear that simple bin-averaging will not suffice in this case, and methods that can reconstruct the mass distribution without resorting to bin-averaging \citep[see, e.g. ][, in prep]{lanusse15} will be preferable. This represents a natural extension to the GLIMPSE approach, and this development is ongoing.

Lastly, we point out that the experiment presented
  here was not blind, and the detection rates for clusters presented
  here do not represent the probability of detecting a cluster of a
  given mass and redshift in a weak lensign survey. Rather, they
  represent the probability of detecting a given cluster given that
  the cluster actually exists (i.e. they give the selection function
  for each reconstruction method). To extend these results to predict
  the number of clusters we expect to detect in a given survey, one
  would need to convolve the cluster selection functions presented
  here with the cluster mass function, which gives the probability of
  the existence of a cluster of a given mass at a given redshift.

Nonetheless, mass mapping remains in our view a
  potentially very useful tool for weak lensing cosmology, and 
GLIMPSE clearly offers several distinct advantages over 2D weak
lensing reconstruction methods. Moreover, it provides a cluster
detection method complementary to optical studies. There is much work
still to be done, but GLIMPSE is clearly a very promising tool for
constraining the non-Gaussian part of cosmological density field and
probing the evolution of nonlinear structures in the Universe.

\section{acknowledgments}
The authors would like to thank Filipe Abdalla and Paniez Paykari for useful
discussions. This work is supported by the European Research Council
grant SparseAstro (ERC-228261). We also gratefully acknowledge the thoughtful and insightful comments provided by the anonymous referee, which have substantially improved the quality of this manuscript. 

\bibliography{mn-jour,refs}

\appendix
\section{Summary of results}

The table below summarises the results of this study as presented in \S~\ref{subsec:results}. Each simulated cluster field is denoted a field number, given in the first column of the table. Columns 2 and 3 give the mass and redshift of the cluster in the field. The four left-most columns give the fraction of reconstructions out of 1000 realisations of a noisy shear field in which the cluster was detected using MRLens with a denoising threshold of $3\sigma$, $3.5\sigma$ and $4\sigma$, and using GLIMPSE with a denoising threshold of $4\sigma$. 
\begin{center}
\small
\tablefirsthead{%
\hline
\multicolumn{1}{|c}{ } & \multicolumn{1}{| l}{Mass } & \multicolumn{1}{ c |}{ } & \multicolumn{3}{| l}{MRLens } & \multicolumn{1}{| l |}{ GLIMPSE} \\
\multicolumn{1}{| l}{Field} & \multicolumn{1}{| l}{($h^{-1}M_\odot$)} & \multicolumn{1}{| c}{z} & \multicolumn{1}{| l}{$3\sigma$} & \multicolumn{1}{| l}{$3.5\sigma$} &  \multicolumn{1}{| l}{$4\sigma$} & \multicolumn{1}{| l |}{$4\sigma$} \\
 \hline}
 \tablehead{%
 \hline
  & Mass  & & MRLens & & & GLIMPSE  \\
Field & ($h^{-1}M_\odot$) & z & $3\sigma$ & $3.5\sigma$ &  $4\sigma$ & $4\sigma$ \\
 \hline}
\tabletail{%
\hline
}
\tablelasttail{\hline}
\begin{supertabular}{| c | l l l l l | l |}
 1 & $3\times10^{13}$ & 0.05 & 0.338 & 0.161& 0.051& 0.134 \\
 2 & $4\times10^{13}$ & 0.05 & 0.431& 0.252& 0.123& 0.221 \\
 3 & $5\times10^{13}$ & 0.05 & 0.579 & 0.383& 0.214& 0.320 \\
 4 & $6\times10^{13}$ & 0.05 & 0.638& 0.447& 0.262& 0.417\\
 5 & $7\times10^{13}$ & 0.05 & 0.738& 0.560& 0.350& 0.520 \\
 6 & $8\times10^{13}$ & 0.05 & 0.803& 0.664& 0.466& 0.610\\
 7 & $9\times10^{13}$ & 0.05 & 0.830& 0.683& 0.507& 0.690\\
 8 & $1\times10^{14}$ & 0.05 & 0.901& 0.795& 0.634& 0.756\\
 9 & $2\times10^{14}$ & 0.05 & 0.991& 0.979& 0.949& 0.988\\
 10 & $3\times10^{14}$ & 0.05 & 0.996 & 0.997& 0.995& 1.000\\
 11 & $4\times10^{14}$ & 0.05 & 0.999& 0.999& 0.999& 1.000\\
 12 & $5\times10^{14}$ & 0.05 & 0.999& 0.998& 0.997& 1.000\\
 13 & $6\times10^{14}$ & 0.05 & 1.000& 1.000& 1.000& 1.000\\
 14 & $7\times10^{14}$ & 0.05 & 1.000& 1.000& 1.000& 1.000\\
 15 & $3\times10^{13}$ & 0.15 & 0.224& 0.083& 0.024& 0.082\\
 16 & $4\times10^{13}$ & 0.15 & 0.338& 0.156& 0.061& 0.149\\
 17 & $5\times10^{13}$ & 0.15 & 0.465& 0.265& 0.111& 0.231\\
 18 & $6\times10^{13}$ & 0.15 & 0.569& 0.351& 0.165& 0.335\\
 19 & $7\times10^{13}$ & 0.15 & 0.671& 0.459& 0.248& 0.442\\
 20 & $8\times10^{13}$ & 0.15 & 0.759& 0.555& 0.346& 0.545\\
 21 & $9\times10^{13}$ & 0.15 & 0.845& 0.658& 0.459& 0.641\\
 22 & $1\times10^{14}$ & 0.15 & 0.909& 0.762& 0.558& 0.732\\
 23 & $2\times10^{14}$ & 0.15 & 0.999& 0.996& 0.981& 0.996\\
 24 & $3\times10^{14}$ & 0.15 & 1.000& 1.000& 1.000& 1.000\\
 25 & $4\times10^{14}$ & 0.15 & 1.000& 1.000& 1.000& 1.000\\
 26 & $5\times10^{14}$ & 0.15 & 1.000 & 1.000& 1.000& 1.000\\
 27 & $6\times10^{14}$ & 0.15 & 1.000& 1.000& 1.000& 1.000\\
 28 & $7\times10^{14}$ & 0.15 & 1.000& 1.000& 1.000& 1.000\\
 29 & $6\times10^{13}$ & 0.25 & 0.361& 0.198& 0.069& 0.192\\
 30 & $7\times10^{13}$ & 0.25 & 0.498& 0.270& 0.117& 0.267\\
 31 & $8\times10^{13}$ & 0.25 & 0.586& 0.351& 0.158& 0.352\\
 32 & $9\times10^{13}$ & 0.25 & 0.636& 0.421& 0.220& 0.437\\
 33 & $1\times10^{14}$ & 0.25 & 0.707& 0.502& 0.290& 0.523\\
 34 & $2\times10^{14}$ & 0.25 & 0.998& 0.967& 0.905& 0.975\\
 35 & $3\times10^{14}$ & 0.25 & 1.000& 1.000& 0.995& 1.000\\
 36 & $4\times10^{14}$ & 0.25 & 1.000& 1.000& 1.000& 1.000 \\
 37 & $5\times10^{14}$ & 0.25 & 1.000& 1.000& 1.000& 1.000\\
 38 & $6\times10^{14}$ & 0.25 & 1.000& 1.000& 1.000& 1.000\\
 39 & $7\times10^{14}$ & 0.25 & 1.000& 1.000& 1.000& 1.000\\
 40 & $8\times10^{14}$ & 0.25 & 1.000& 1.000& 1.000& 1.000\\
 41 & $9\times10^{14}$ & 0.25 & 1.000& 1.000& 1.000& 1.000\\
 42 & $1\times10^{15}$ & 0.25 & 1.000& 1.000& 1.000& 1.000\\
 43 & $6\times10^{13}$ & 0.35 & 0.207& 0.098& 0.044& 0.110\\
 44 & $7\times10^{13}$ & 0.35 & 0.296& 0.129& 0.039& 0.155\\
 45 & $8\times10^{13}$ & 0.35 & 0.373& 0.201& 0.080& 0.210\\
 46 & $9\times10^{13}$ & 0.35 & 0.448& 0.254& 0.099& 0.267\\
 47 & $1\times10^{14}$ & 0.35 & 0.513& 0.289& 0.120& 0.333\\
 48 & $2\times10^{14}$ & 0.35 & 0.944& 0.831& 0.664& 0.887\\
 49 & $3\times10^{14}$ & 0.35 & 0.999& 0.988& 0.955& 0.995\\
 50 & $4\times10^{14}$ & 0.35 & 1.000& 1.000& 0.999& 1.000\\
 51 & $5\times10^{14}$ & 0.35 & 1.000& 1.000& 1.000& 1.000\\
 52 & $6\times10^{14}$ & 0.35 & 1.000& 1.000& 1.000& 1.000\\
 53 & $7\times10^{14}$ & 0.35 & 1.000& 1.000& 1.000& 1.000\\
 54 & $8\times10^{14}$ & 0.35 & 1.000& 1.000& 1.000& 1.000\\
 55 & $9\times10^{14}$ & 0.35 & 1.000& 1.000& 1.000& 1.000\\
 56 & $1\times10^{15}$ & 0.35 & 1.000& 1.000& 1.000& 1.000\\
 57 & $9\times10^{13}$ & 0.45 & 0.250& 0.117& 0.035& 0.157\\
 58 & $1\times10^{14}$ & 0.45 & 0.300& 0.137& 0.044& 0.199\\
 59 & $2\times10^{14}$ & 0.45 & 0.772& 0.574& 0.361& 0.710\\
 60 & $3\times10^{14}$ & 0.45 & 0.964& 0.889& 0.756& 0.959\\
 61 & $4\times10^{14}$ & 0.45 & 1.000& 0.988& 0.947& 0.997\\
 62 & $5\times10^{14}$ & 0.45 & 0.997& 0.997& 0.990& 1.000\\
 63 & $6\times10^{14}$ & 0.45 & 1.000& 1.000& 0.999& 1.000\\
 64 & $7\times10^{14}$ & 0.45 & 1.000& 1.000& 1.000& 1.000\\
 65 & $8\times10^{14}$ & 0.45 & 1.000& 1.000& 1.000& 1.000\\
 66 & $9\times10^{14}$ & 0.45 & 1.000& 1.000& 1.000& 1.000\\
 67 & $1\times10^{14}$ & 0.55 & 0.210& 0.094& 0.020& 0.115\\
 68 & $2\times10^{14}$ & 0.55 & 0.556& 0.322& 0.135& 0.491\\
 69 & $3\times10^{14}$ & 0.55 & 0.836& 0.681& 0.432& 0.834\\
 70 & $4\times10^{14}$ & 0.55 & 0.963& 0.896& 0.746& 0.969\\
 71 & $5\times10^{14}$ & 0.55 & 0.997& 0.975& 0.912& 0.997\\
 72 & $6\times10^{14}$ & 0.55 & 0.999& 0.991& 0.967& 1.000\\
 73 & $7\times10^{14}$ & 0.55 & 1.000& 1.000& 0.993& 1.000\\
 74 & $8\times10^{14}$ & 0.55 & 1.000& 1.000& 1.000& 1.000\\
 75 & $9\times10^{14}$ & 0.55 & 1.000&1.000& 1.000& 1.000\\
 76 & $1\times10^{15}$ & 0.55 & 1.000& 1.000& 1.000& 1.000\\
 77 & $1\times10^{14}$ & 0.65 & 0.103& 0.037& 0.009& 0.067\\
 78 & $2\times10^{14}$ & 0.65 & 0.355& 0.186& 0.051& 0.309\\
 79 & $3\times10^{14}$ & 0.65 & 0.589& 0.375& 0.186& 0.634\\
 80 & $4\times10^{14}$ & 0.65 & 0.806& 0.620& 0.409& 0.862\\
 81 & $5\times10^{14}$ & 0.65 & 0.928& 0.806& 0.616& 0.963\\
 82 & $6\times10^{14}$ & 0.65 & 0.971& 0.921& 0.794& 0.994\\
 83 & $7\times10^{14}$ & 0.65 & 0.986& 0.968& 0.910& 0.999\\
 84 & $8\times10^{14}$ & 0.65 & 1.000& 0.988& 0.956& 1.000\\
 85 & $9\times10^{14}$ & 0.65 & 1.000& 1.000& 0.990& 1.000\\
 86 & $1\times10^{15}$ & 0.65 & 1.000& 0.998& 0.996& 1.000\\
 87 & $1\times10^{14}$ & 0.75 & 0.073& 0.015& 0.005& 0.041\\
 88 & $2\times10^{14}$ & 0.75 & 0.160& 0.065& 0.016& 0.178\\
 89 & $3\times10^{14}$ & 0.75 & 0.368& 0.189& 0.080& 0.415\\
 90 & $4\times10^{14}$ & 0.75 & 0.577& 0.344& 0.187& 0.657\\
 91 & $5\times10^{14}$ & 0.75 & 0.691& 0.497& 0.283& 0.838\\
 92 & $6\times10^{14}$ & 0.75 & 0.825& 0.651& 0.430& 0.937\\
 93 & $7\times10^{14}$ & 0.75 & 0.905& 0.783& 0.596& 0.979\\
 94 & $8\times10^{14}$ & 0.75 & 0.959& 0.888& 0.722& 0.995\\
 95 & $9\times10^{14}$ & 0.75 & 0.980& 0.934& 0.813& 0.999\\
 96 & $1\times10^{15}$ & 0.75 & 0.986& 0.962& 0.888& 1.000\\
 \end{supertabular}
 \end{center}

\label{lastpage}

\end{document}